\RequirePackage{fix-cm}
\documentclass[fleqn,usenatbib]{mnras}

\usepackage{newtxtext,newtxmath}

\usepackage[T1]{fontenc}

\DeclareRobustCommand{\VAN}[3]{#2}
\let\VANthebibliography\thebibliography
\def\thebibliography{\DeclareRobustCommand{\VAN}[3]{##3}\VANthebibliography}

\usepackage{multirow}
\usepackage{hyperref}
\usepackage{longtable}
\usepackage{pdflscape}
\usepackage{subcaption}

\usepackage{graphicx}	% Including figure files
\usepackage{amsmath}	% Advanced maths commands

\usepackage[dvipsnames]{xcolor}

\title[Early thin-disc assembly]{Early thin-disc assembly revealed by JWST edge-on galaxies}

\author[M. van Asselt et al.]{
Marloes van Asselt,$^{1}$
Francesca Rizzo,$^{1}$\thanks{E-mail: rizzo@astro.rug.nl}
Luca Di Mascolo$^{1}$
\\
$^{1}$Kapteyn Astronomical Institute, University of Groningen, Landleven 12, 9747 AD, Groningen, The Netherlands
}

\date{Accepted XXX. Received YYY; in original form ZZZ}

\pubyear{\the\year{}}

\begin{document}
\label{firstpage}
\pagerange{\pageref{firstpage}--\pageref{lastpage}}
\maketitle

% Abstract of the paper
\begin{abstract}
The vertical structure of stellar discs provides key constraints on their formation and evolution. Nearby spirals, including the Milky Way, host thin and thick components that may arise either from an early turbulent phase or from the subsequent dynamical heating of an initially thin disc; measuring disc thickness across cosmic time therefore offers a direct test of these scenarios. We present a new methodology to measure the thickness of edge-on galaxies that explicitly accounts for departures from perfectly edge-on orientations by fitting a full three-dimensional model with forward modelling. This improves on traditional approaches that assume an inclination of $90^\circ$ and can bias thicknesses high. Applying the method to \textit{JWST} imaging of galaxies at $1<z<3$ with stellar masses $\gtrsim 10^9 M_\odot$ from four major surveys, we measure a median scale height of $z_0 = 0.25\pm0.14$~kpc and a median ratio $h_r/z_0=8.4\pm3.7$. These values are consistent with the Milky Way and local thin discs, and indicate scale heights $\sim 1.6$ times smaller than those inferred for local galaxies from single-disc fits. This result implies that thin discs are already present at $z\sim3$. We further show that a thick disc contributing 10\% of the thin-disc luminosity would be detectable in the data considered in this work, implying that any thick disc present must be fainter and favouring a scenario in which thick discs build up progressively through dynamical heating at $z<1$.

\end{abstract}

% Select between one and six entries from the list of approved keywords.
% Don't make up new ones.

%%%%%%%%%%%%%%%%%%%%%%%%%%%%%%%%%%%%%%%%%%%%%%%%%%

%%%%%%%%%%%%%%%%% BODY OF PAPER %%%%%%%%%%%%%%%%%%

\section{Introduction}
\label{Sec:intro}

The stellar discs of the Milky Way has a bimodal structure, typically described in terms of a thin and a thick disc component \citep[e.g.,][]{Gilmore_reid_1983, Bland_Hawthorn_2016, Gallart_2025}. The thin disc is dynamically cold and hosts stellar populations confined within a small vertical scale height of about 300 pc \citep{Bland_Hawthorn_2016}. In contrast, the thick disc is dynamically hot, and composed predominantly of stars extending up to approximately 1 kpc above the Galactic plane \citep{1977_Wielen, 2003_Bensby, 2013_adibekyan, 2014_Recio-blanco, Tkachenko_2025}. Similar dual-disc structures have been identified in external spiral galaxies \citep{1996_degrijs, seth2005verticalstructurestarsedgeon, 2006_yoachim, 2019_Pinna, 2025_xu}, suggesting that the coexistence of thin and thick discs is a common outcome of galaxy evolution \citep[see][for a recent review on disc thickness]{JOG_2026}.

In the Milky Way, where stellar ages and detailed chemical abundances can be measured, it has long been recognized that the distinction between the thin and thick discs extends beyond structure and kinematics to include both chemistry and age. At a given metallicity, thick-disc stars are typically more $\alpha$-enhanced (higher $[\alpha/\mathrm{Fe}]$) than their thin-disc counterparts \citep[e.g.,][]{Fuhrmann_2011, Haywood_2013, Anders_2018, Silva_2018}. This chemical bimodality is mirrored in stellar ages: the high-$\alpha$ population is dominated by old stars, with a mean age of $\sim 11$~Gyr, while, the low-$\alpha$ population is significantly younger on average, with mean ages of $\sim 5$--6~Gyr \citep[e.g.,][]{Miglio_2021, 
Lagarde_2021, Queiroz_2023}. It is worth noting, however, that this two-component description is itself a simplification. Studies of mono-abundance populations have shown that disc structural properties vary continuously with chemistry, with no clear discontinuity between a thin and a thick component \citep{Bovy_2012}. Furthermore, the geometrically defined thin and thick discs do not always bear a one-to-one correspondence with the chemically defined low-$\alpha$
 and high-$\alpha$ populations \citep[e.g.,][]{Hayden_2017, Vincenzo_2021}. Stars selected geometrically can span a range of chemical abundances, and conversely, chemical selection can recover populations with overlapping spatial distributions.

While the chemical and age trends of the thin and thick discs are well established, at least in the Milky Way, both the timing of disc formation and the physical origin of the thin/thick disc dichotomy have been the subject of intense debate in recent years. Studies of metal-poor stars increasingly support a scenario in which the Galactic disc assembled very early \citep[e.g.,][]{Sestito_2019, DiMatteo_2020, Viswanathan_2025}, with recent work suggesting that disc formation began at $\gtrsim 12$~Gyr ago \citep[e.g.,][]{Hou_2025, Xiang_2025, Orkney_2026}. The age of the thin disc, however, remains uncertain. Some studies argue for the presence of an old thin-disc population \citep[$>11$~Gyr,][]{Beraldo_2021, Nepal_2024, Vorbolato_2025}, implying a period of co-formation of thin and thick components. Other works instead support an evolutionary sequence in which the thick disc formed first, while the thin disc emerged more recently; in this framework, it remains unclear whether the earliest disc was intrinsically thin or already thick \citep{Chandra_2024, Viswanathan_2025, Xiang_2025}. More broadly, it is uncertain whether the Milky Way's disc properties are representative of the general disc-galaxy population. Comparisons with cosmological simulations suggest that aspects of the Milky Way's formation history, such as its early disc assembly and relatively quiescent recent merger history, may not be typical of present-day disc galaxies \citep[e.g.,][]{Evans_2020, Dillamore_2024, Semenov_2024, McCluskey_2025}.

Several formation pathways have been proposed for explaining the formation of the thin and thick discs, broadly falling into two main categories. In the first scenario, a galaxy initially forms a disc that is subsequently dynamically heated to form a thick disc \citep{2021meng, Dodge_2023}. This heating can occur via rapid external processes such as mergers or satellite accretion \citep{1996_robin, 2003_Abadi}, or through slower internal mechanisms, including gravitational interactions between stars and non-axisymmetric structures such as spiral arms, clumps, or bars \citep{1993_Quinn, 1996_Walker, villalobos_2008, 2011_Qu, 2009_Bournaud, 2011_comeron, 2016_Aumer, 2016_Grand}, as well as the diffusion of stellar orbits due to encounters with massive star clusters \citep{2002_Kroupa, 2011_Assmann}. However, there is still no consensus on whether dynamical heating from these processes alone is sufficient to reproduce the properties of the thick discs \citep[e.g.,][]{McCluskey_2025}.

In the second scenario, the thick disc forms directly from a turbulent, gas-rich phase of early disc evolution, producing stars with high intrinsic velocity dispersions \citep{1992_Burkert, brook_2004, 2009_Bournaud, 2013_bird, 2017_Leaman, 2020_Grand, 2021_bird, 2021_Yu, 2023_Yu}. A subsequent, more quiescent phase of star formation gives rise to the thin disc. This interpretation is supported by observations showing that star-forming galaxies at high redshift have significantly higher gas velocity dispersions than local spiral galaxies \citep[e.g.,][]{2015_Wisnioski, Simons_2017, 2018_Johnson, Danhaive_2025}. These measurements are typically based on warm ionized gas tracers, such as H$\alpha$ emission, under the assumption that H$\alpha$ kinematics reliably trace the motions of the molecular gas. However, this assumption is known to break down in nearby galaxies: velocity dispersions inferred from H$\alpha$ and other ionized-gas tracers are typically higher than those measured from molecular or atomic gas tracers \citep[e.g.,][]{Levy_2018, Girard_2021}. Recent work further suggests that, even at high redshift, H$\alpha$-based dispersions may similarly overestimate the turbulent motions of the cold gas \citep[e.g.,][]{Kohandel_2024, 2024_rizzo, Parlanti_2024, Jones_2025}. Observational confirmation of this effect, however, remains largely limited to massive, highly star-forming systems.

A powerful way to constrain the formation and evolution of galactic discs is to study how their kinematics, as measured using cold gas tracers, change across cosmic time. However, the sensitivity limits of current sub-millimetre interferometers make it extremely challenging to directly measure gas kinematics at $z > 1$, which is currently feasible only for highly star-forming galaxies with stellar masses $\gtrsim 10^{10.5}~\mathrm{M_{\odot}}$ \citep[see compilation in][]{2024_rizzo}. An alternative approach is to study the vertical structure of stellar discs over cosmic time, which can provide indirect yet valuable insights into their assembly and dynamical evolution. The two scenarios described above predict distinct redshift trends in disc thickness: the heating scenario implies that discs were thinner at earlier epochs and became progressively thicker toward the present day, whereas the in-situ formation scenario predicts that discs were already thick at high redshift (i.e., $z > 1$).

Before the launch of \textit{JWST}, some studies attempted to measure the thicknesses of high-redshift galaxies using samples of edge-on systems up to $z \sim 2$ \citep{Elmegreen_2017, 2023-Hamilton-campos}. These works found median scale heights comparable to the Milky Way thick disc but smaller than the median scale height of local discs ($\sim 1.5~\mathrm{kpc}$). They concluded that discs as thick as the Milky Way’s thick disc were already in place at early times, but explaining the full population of present-day disc galaxies would imply that stellar discs must have experienced additional physical thickening after formation \citep{2023-Hamilton-campos}. The advent of \textit{JWST} has revolutionized such studies by providing near-infrared imaging with significantly improved angular resolution compared to the Hubble Space Telescope (\textit{HST}) \citep{NIRcam}. Moreover, \textit{JWST}/NIRCam filters extend up to $4.4~\mathrm{\mu m}$, enabling rest-frame near-infrared observations of galaxies out to $z \sim 3$, where dust attenuation effects are significantly reduced \citep{introductiongalaxyformationevolution}.

To date, two studies have measured the vertical thickness of high-redshift galaxies using \textit{JWST} data \citep{2024_lian, 2024Tsukui}. \citet{2024_lian} analysed 191 galaxies up to $z \sim 5$ and found that high-redshift systems ($z \geq 1.5$) have larger scale heights than their lower-redshift counterparts, consistent with discs forming already thick. Similarly, \citet{2024Tsukui} modelled 132 galaxies at $z \sim 0.1-3$ and found no significant correlation between disc scale height and redshift. They interpreted this as evidence that thick discs form early, with thin discs assembling later, while the thick component continues to grow as galaxies evolve. Nevertheless, current measurements of stellar disc thickness may be systematically biased toward higher values due to limitations in the methodologies employed. Most studies investigating the vertical structure of stellar discs in external galaxies have relied on one-dimensional (1D) analyses \citep{2006_yoachim, Elmegreen_2017, 2023-Hamilton-campos, 2024_lian}. In such methods, the vertical surface brightness profile is extracted along a line perpendicular to the disc and fitted with a parametric model, often simplifying or neglecting the effects of the PSF. Some studies, in fact, omit PSF convolution entirely \citep[e.g.,][]{2006_yoachim}, while others include it by convolving a line of infinitesimal thickness with the PSF \citep[e.g.,][]{Elmegreen_2017, 2023-Hamilton-campos, 2024_lian}. Both treatments can be suboptimal, potentially leading to artificially inflated estimates of disc thickness. \citet{2024Tsukui} recently introduced a more sophisticated approach, employing two-dimensional (2D) forward modelling of the surface brightness distribution. In this method, a parametric model of an edge-on galaxy is convolved directly with the PSF and compared to the observed image, thereby providing a more realistic treatment of instrumental effects. Still, their analysis assumes that galaxies are perfectly edge-on, without accounting for the inclination angle of the galaxy ($i$) relative to the line of sight. This simplification can introduce systematic overestimates of the vertical scale height ($z_0$) when the true inclination deviates from 90$^\circ$ \citep{1997_degrijs}.

In this paper, we present a new methodology to measure the vertical thickness of edge-on galaxy discs using a forward-modelling technique that simultaneously fits for galaxy inclination and fully accounts for PSF convolution. We apply this model to \textit{JWST} observations of galaxies at $z > 1$ to investigate the emergence of the thin and thick discs. The paper is structured as follows. In Section~\ref{sec:data}, we describe the sample selection and the corresponding data. Section~\ref{sec:methodology} introduces the model developed to measure the disc thickness. In Sections~\ref{sec:results} and~\ref{sec:comparison}, we present the results and compare them with measurements from the literature. In Section~\ref{sec:emergence}, we discuss the implications of our findings for formation scenarios of thin and thick disc components. In Section~\ref{sec:discussion}, we further discuss the implications of our results for high-$z$ galaxy kinematics and the formation history of the Milky Way. Finally, we summarise the main results in Section~\ref{sec:conclusion}. Throughout this work, we adopt the \citet{2016_planck} cosmology, with $H_0 = 67.8\,\mathrm{km\,s^{-1}\,Mpc^{-1}}$, 
$\Omega_{\mathrm{m}} = 0.308$, and
$\Omega_{\Lambda} = 0.692$, 
and $\Omega_{\mathrm{b}} = 0.0486$. Unless stated otherwise, all the best-fit values of the model parameters are computed as the median of the corresponding marginal posterior probability distribution. The associated lower and upper uncertainties are obtained as difference between these median values and the 16\textsuperscript{th} and 84\textsuperscript{th} quantiles, respectively.

\section{Data}
\label{sec:data}
As is standard for studies aiming to measure the vertical thickness of stellar discs \citep[e.g.,][]{2006_yoachim, 2023-Hamilton-campos, 2024Tsukui}, our targets are edge-on galaxies, where the vertical structure can be directly observed and modelled. In Section~\ref{sec:cat}, we describe the catalogue used to identify edge-on systems and outline the selection criteria adopted to construct the final sample. Section~\ref{sec:data_prep} presents the preprocessing steps applied to the imaging data prior to the analysis.

\subsection{Sample selection}
\label{sec:cat}
To find edge-on galaxy candidates, we use the morphological catalogue from the DAWN \textit{JWST} Archive \citep[DJA,][]{genin_2024}. This catalogue is based on the surface brightness fitting of NIRCAM images from four of the major extragalactic \textit{JWST} surveys. This includes the COSMOS and UDS fields from PRIMER \citep{donnan_JWST_2024}, GOODS North and South from FRESCO \citep{oesch_JWST_2023} and EGS from CEERS \citep{finkelstein_cosmic_2025}. Together, these fields cover nearly 390 square arcminutes of the sky and have a $5\sigma$ depth of $\gtrsim 28$ mag in the filters F227W, F356W and F444W.

To model the surface brightness of the galaxies, \citet{genin_2024} fitted the 2D images of the galaxies in all available filters assuming two components: a disc and a bulge described by exponential and Sérsic profiles \citep{1963_sersic}, respectively. The Sérsic profile is defined by:
\begin{equation}
\label{eq:sersic}
    I(R)= I_\mathrm{e} \exp \left\{ -b_n \left[ \left(\frac{R}{R_\mathrm{e}} \right)^{1/n}-1 \right] \right\},
\end{equation}
where $R_\mathrm{e}$ is the effective radius, $I_\mathrm{e}$ is the surface brightness at the effective radius, $n$ is the Sersic index, $b_n$ is a parameter that depends on $n$ \citep{Ciotti_1999}. For $n=1$, the Sérsic profile becomes the exponential profile. 

\citet{genin_2024} performed the fit using \texttt{SourceXtractor++} \citep{2020_Bertin,2022_Kummel} and provide a public catalogue which includes the effective radius of the disc and its axis ratio $q$, which is defined as the ratio between the semi-minor and the semi-major axes. In addition to the parameters defining the bulge and the disc components, the catalogue provides parameters obtained from Spectral Energy Distribution (SED) fitting using \texttt{EAZY} \citep{2008_brammer}, such as the stellar mass ($M_\star$), the photometric redshift ($z_{\mathrm{phot}}$) and the star formation rate (SFR).

To identify $z > 1$ edge-on galaxy candidates and exclude targets that can result in unreliable parameters, we applied the following selection criteria using the information and flags provided in the DJA morphological catalogue:
\begin{enumerate} 
    \item We retain only objects with morphological flag = 2 (successful fit and science-ready data), excluding sources with flags indicating artifacts, stars, or low signal-to-noise (S/N $< 3$).
    \item Observable at restframe $\lambda_{\mathrm{rest}}= 1.2~\mathrm{\mu m}$. We aim to observe similar rest-frame emission across all redshifts; therefore, we adopt a reference wavelength of $\lambda_{\mathrm{rest}} = 1.2~\mathrm{\mu m}$. This choice minimizes the effects of dust attenuation while allowing us to probe a broad redshift range. However, it represents a conservative choice, as the near-infrared emission at this wavelength is more sensitive to older stellar populations that may have already undergone dynamical heating. 
    \item $M_\star \geq 10^8~\mathrm{M_{\odot}}$. This lower limit is due to the fact that we expect galaxies below this limit to be unresolved at $1.2~\mathrm{\mu m}$ based on the size-mass relation \citep{2014vanderwel,2025_Yang,2025_Miller}.
    \item (Nearly) edge-on.
    We are only interested in galaxies that are edge-on. Following the convention used in previous studies \citep{Elmegreen_2017, 2024Tsukui}, we select galaxies with an axis ratio of $q \leq 0.3$.
    \item Resolvable effective radius. We take into consideration the angular size of the galaxies compared to the PSF FWHM of the \textit{JWST} filters. We set the requirement that the effective radius of the disc needs to be $R_\mathrm{e}\geq 2\times$ FWHM. This is to ensure that the radial size of the galaxy is resolvable.
    \item No neighbouring sources within 2 arcsec. We first exclude any sources which are within 2 arcsec of each other, since we do not want to have any contamination from nearby sources when fitting the data. This can partially be solved by masking sections of the image but problems will still arise when dealing with very bright or overlapping sources.
\end{enumerate}

\begin{figure*}
    \centering
    \includegraphics[width=\linewidth]{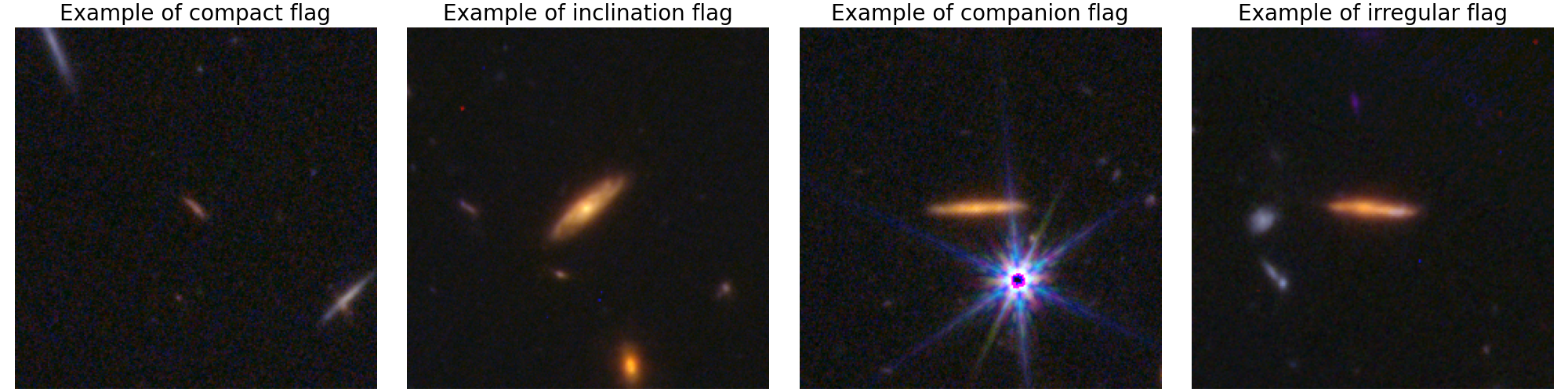}
    \caption{Examples of galaxies excluded from the final sample based on visual inspection. From left to right: compact objects that are too small for reliable fitting; systems that are not fully edge-on; a companion/contaminated case where a nearby or foreground source affects the image and cannot be effectively masked; and irregular galaxies excluded due to strong asymmetry.}
    \label{fig:flag_ex}
\end{figure*}

The application of these selection criteria results in 1024 galaxy candidates. We visually inspect these 1024 galaxies and flag candidates that include objects that are too compact, galaxies with spiral features, indicating that the galaxy is slightly face-on, or source with nearby companions that were not included in the parent catalogue and therefore not filtered out. Examples of galaxies with these flags are given in Fig. \ref{fig:flag_ex}.

\begin{figure*}
    \centering
    \includegraphics[width=\linewidth]{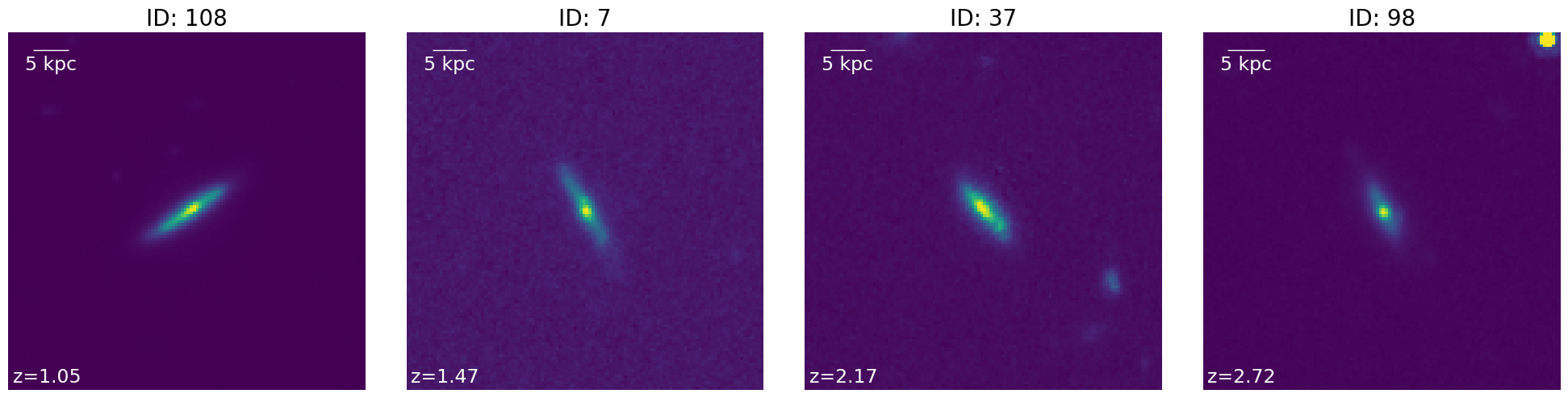}
    \caption{Four representative galaxies included in our final sample.}
    \label{fig:final_sample_ex}
\end{figure*}

After visual inspection, 111 of them do not have any of the aforementioned flags. Since we use an axisymmetric model to fit the thickness of our galaxies (see Section \ref{sec:methodology} for details), we exclude from the final sample galaxies that are asymmetric (e.g., lopsided, or showing warp-like distortions) or clumpy (see right panel, Fig. \ref{fig:flag_ex}). Out of the 111 galaxies, 21 show these asymmetries and/or irregularity and are removed from the final sample which, therefore, consists of 90 galaxies. Fig. \ref{fig:final_sample_ex} shows some examples of galaxies in the final sample.
% and the full sample is shown in Appendix \ref{sec:full_selection}.

\begin{figure}
    \centering
    \includegraphics[width=\linewidth]{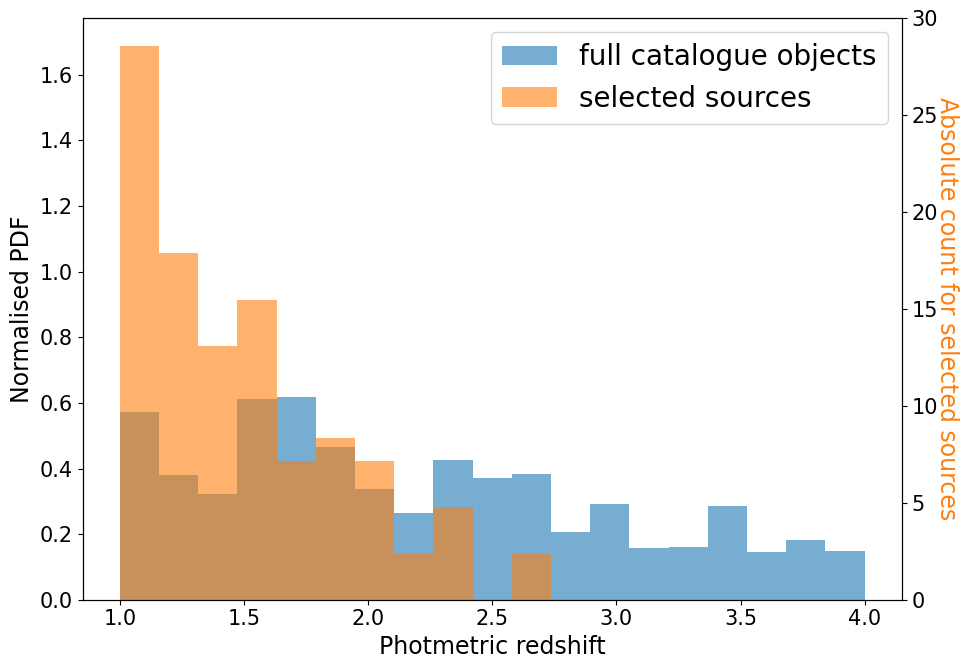}
    \caption{ Normalised distribution of redshift of the 90 galaxies in our sample (orange) compared to the parent sample (blue). The right axis shows the absolute count for the selected sources per redshift bin. The target selection provide a final sample biased towards the lower redshift sources, without any selected sources above z = 3.}
    \label{fig:redshift_distr}
\end{figure}

\begin{figure}
    \centering
    \includegraphics[width=\columnwidth]{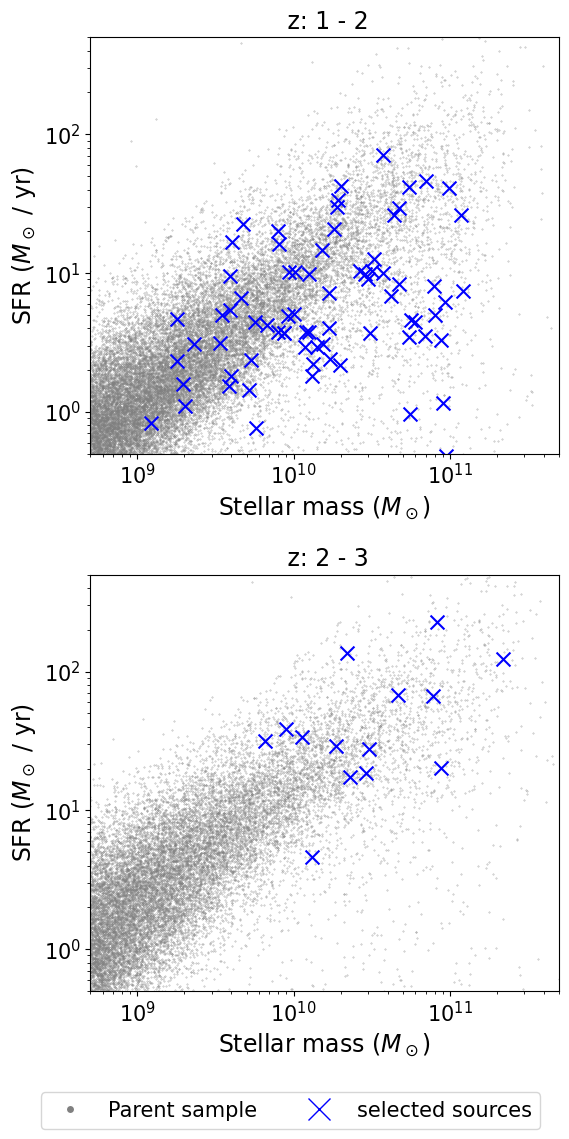}
    \caption{Distribution of SFR vs $M_\star$ per redshift bin of the selected sample (blue crosses) compared to the parent DJA sample (gray circles).}
    \label{fig:mass-sfr-dstr}
\end{figure}

Fig. \ref{fig:redshift_distr} shows the distribution of redshift of the selected galaxies compared to the distribution of the whole dataset of $\sim$ 136,000 galaxies at redshift $z=1-4$ in the DJA catalogue. We notice that there is a bias towards galaxies at $z<2$. This is to be expected, considering that finding resolvable galaxies becomes more difficult as redshift increases. 
In Fig. \ref{fig:mass-sfr-dstr}, we show the SFR vs $M_\star$ of the selected sample with blue cross compared to the DJA catalogue (gray circles). The selected samples follow a similar distribution compared to the parent sample and it can be considered representative of galaxies with stellar masses $\gtrsim 10^9~\mathrm{M_{\odot}}$ at $z = 1 - 2$ and galaxies with stellar masses $\gtrsim 10^{10}~\mathrm{M_{\odot}}$ at $z = 2 - 3$. We note that \citet{2024Tsukui} apply selection criteria broadly similar to ours, resulting in a sample with comparable number statistics and similar distributions in stellar mass and redshift.\\

\subsection{Data preparation}
\label{sec:data_prep}
\subsubsection{Image cutouts}
Our analysis is performed on cutouts centred on the galaxy targets. The angular size of the cutouts was initially set at 3 arcseconds, but was increased for some of the more extended sources up to 6 arcseconds. This angular size was chosen such that all target objects occupy a similar portion of the total Field of View. The filter for each image is decided based on a restframe wavelength of $1.2~\mathrm{\mu m}$. All images have a pixel-scale of 0.05 arcsecond per pixel.

\subsubsection{Signal-to-noise ratio estimation}
\label{sec:RMS}
We quantify the background noise level in each image in terms of their respective Root Mean Square (RMS). To identify the region of pure background, we adopt an iterative sigma clipping approach. For the sigma clipping function, we use the implementation provided by the \texttt{python} package \texttt{photutils}\footnote{https://photutils.readthedocs.io/en/stable/index.html} \citep{photutils}. At every step, the algorithm clips out any sources above a certain significance threshold, based on the standard deviation of the values of all the unmasked pixel. The masked image is then used to compute a new estimate of the standard deviation and apply another step of clipping. Such iteration is then repeated until convergence, i.e., until a stable image is reached or for a maximum of 10 iterations. In our case, we set the threshold to $3\sigma$. The noise RMS is then computed on the final clipped image, assumed to comprise only background signal. Then, to quantify the S/N, we use the same threshold of $3\sigma$ on images with neighbouring sources masked (see section \ref{sec:masks}). This means that we only consider pixels to contain a signal if they are at least 3 times higher than the RMS. Out of these pixels, we take the average value of signal/RMS to be the S/N of the image. Across our sample, the resulting S/N spans 7–77, with a median of 24.

\subsubsection{Masks}
\label{sec:masks}
Some sources in the final sample contain neighbouring sources. This could be either faint compact sources or background/foreground galaxies close to the edges of our cutouts. These sources were not filtered out by the pre-filtering of the data and were not flagged during visual inspection because they are far enough from the target that they can be simply masked out during the fitting. To build the masks, we use the \texttt{SourceFinder} function from the \texttt{photutils} package. This function allows the user to define a minimum number of pixels and a source threshold and returns a segmentation image with the same size as the input data, with all sources labelled with an integer, with zero being the background.
We set the criteria of minimum of 10 pixels per source and the signal threshold is once again $3\sigma$. After obtaining the segmentation map, all sources near the centre (within $15-20$ pixels from the centre) are taken to be part of the target image while others are included in the mask.

\section{Methods}
\label{sec:methodology}
In order to obtain the scale height of the selected targets, we use a forward modelling approach. We generate a three-dimensional (3D) model of a galaxy and we project it onto the plane of the sky where it is compared with the observed galaxy images, after accounting for the convolution with the PSF. In Section \ref{sec:3d}, we will discuss how we build the 3D model; in Section \ref{sec:ins}, we describe the fitting methodology; in Section \ref{sec:mockdata_gen} and \ref{section:dataval}, we discuss the validation of our model using mock data. Finally, in Section \ref{sec:fitting_real}, we describe the assumptions we made to fit the galaxies in our samples.

\subsection{3D Model}
\label{sec:3d}
To construct a 3D model of a galaxy, we assume that the galaxy is axisymmetric and consists of a disc with finite thickness. We adopt both cylindrical ($R, Z$) and Cartesian coordinates ($X, Y, Z$), where $X$ and $Y$ are the coordinates in the plane of the galaxy (i.e., $\sqrt{X^2+Y^2} = R$) and $Z$ being perpendicular to the plane. Under these assumptions, the galaxy emissivity is modelled as
\begin{equation}
    \label{eq:brightnessprofile}
    B(R,Z)=I_0 \exp\left(-\frac{R}{h_r}\right)\mathrm{sech}^2\left(\frac{Z}{2\,z_0}\right)
\end{equation}
The exponential function is used to model the emission of the disc in the galactic plane $(Z = 0)$. For the vertical distribution, we assume a squared hyperbolic secant profile. This functional form was originally derived for a self-gravitating, isothermal system in vertical hydrostatic equilibrium by \citep{Spitzer_1942} and \citep{Camm_1950}. Under these conditions, the solution of the joint Poisson-hydrostatic balance equation yields a $\mathrm{sech}^2$ vertical density distribution. Observationally, this profile has been shown to provide a good description of the vertical structure of edge-on disc galaxies \citep{1981_kruit}.
The sech$^2$ function is characterised by the following approximations:
\begin{equation}
    Z/z_0' \ll 1: \mathrm{sech}^2(Z/z_0')=\text{exp}(-Z^2/z'^2_0)
    \label{eq:sech_smallz}
\end{equation}
\begin{equation}
    Z/z'_0 \gg 1: \mathrm{sech}^2(Z/z'_0)=4 \text{ exp}(-2Z/z'_0)
    \label{eq:sech_largez}
\end{equation}
The sech$^2$ profile shows Gaussian behaviour near the galactic plane (Z=0) and changes into an exponential profile at larger Z. Here we see that this exponential has a scaleheight of $z'_0/2$. This is also how we chose to define the scale height in the brightness profile given in equation (\ref{eq:brightnessprofile}), $z_0=z'_0/2$. The definition of the scale height in the literature depends on the adopted functional form. Throughout this work, we use $z_0$ as defined in equation (\ref{eq:brightnessprofile}). This definition is taken to be equivalent to the
$z_0$ used in the exponential profile, $\mathrm{exp}(-Z/z_0)$ \citep[values found using $\mathrm{sech}^2(Z/2*z_0)$ are 10\% smaller compared to an exponential profile][]{Bland_Hawthorn_2016}, and corresponds to half the value of $z_0$ in studies that adopt the $\mathrm{sech}^2(Z/z_0)$ profile.

\begin{figure*}
    \centering
    \includegraphics[width=\linewidth]{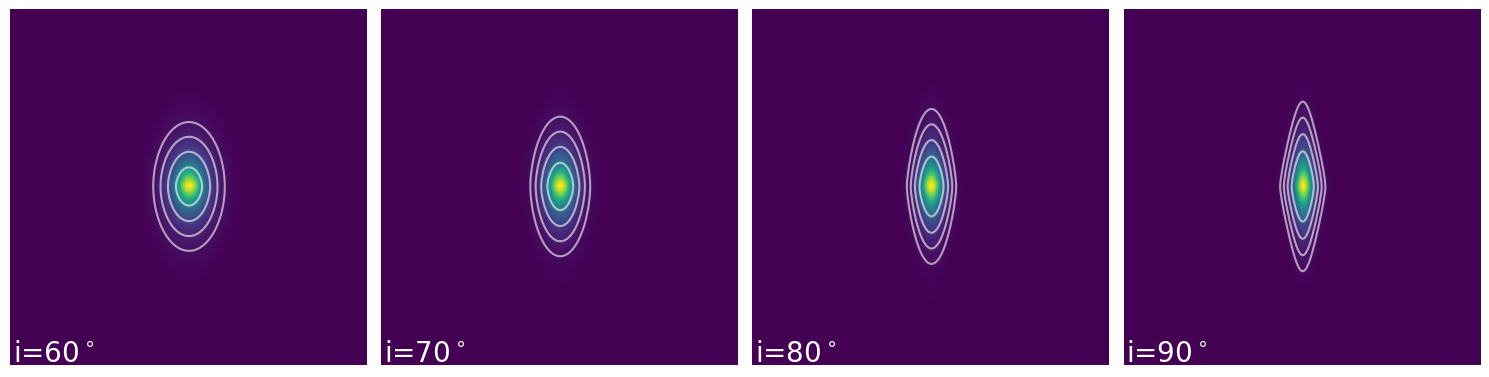}
    \caption{The different panels show mock observations of the same disc and its isophotes (white contours) viewed at different inclinations. }
    \label{fig:Inclination}
\end{figure*}

In this work, we explicitly account for variations in inclination, since it is unlikely that all galaxies are perfectly edge-on. Even small departures from $i = 90^\circ$ can make a disc appear thicker than it truly is. Fig.~\ref{fig:Inclination} illustrates this effect with mock observations of the same finite-thickness disc viewed at different inclinations, showing how a slight tilt can mimic increased vertical thickness.\\ 

Once the 3D model is defined on the galactic cylindrical reference frame, it is transformed via three-dimensional rotation to match the sky-plane coordinate system. This is defined by right ascension (RA), declination (Dec), and the line of sight (LoS). The transformation from the model frame to (RA, Dec, LoS) is implemented via two rotation matrices, $\mathbf{R}_1$ and $\mathbf{R}_2$, such that:

\begin{equation}
    \label{eq:rotation}
    \begin{pmatrix}
         \mathrm{RA}\\ \mathrm{Dec}\\ \mathrm{LoS}
    \end{pmatrix} = \textbf{R}_2 \cdot \textbf{R}_1\cdot
    \begin{pmatrix}
        X \\ Y\\Z
    \end{pmatrix}
\end{equation}
Here, ${\textbf{R}_1}$ is the rotation matrix defining the rotation of the plane of the galaxy with respect to the LoS. This rotation is defined by the inclination angle ($i$) that is the angle between the LoS and the axis Z. The matrix ${\textbf{R}_1}$ is therefore defined as:
\begin{equation}
    \textbf{R}_1 =
    \begin{pmatrix}
         1 &0&0\\
         0& \mathrm{cos}(i) & - \mathrm{sin}(i)\\ 
         0& \mathrm{sin}(i) & \mathrm{cos}(i)
    \end{pmatrix} 
\end{equation}
The matrix $\textbf{R}_2$ defines the rotation on the plane of the sky and it is defined by the position angle $\theta$ defining the orientation of the projected major axis, measured from north to east.
\begin{equation}
    \textbf{R}_2=
    \begin{pmatrix}
        \mathrm{sin}(\theta) &0& \mathrm{cos}(\theta)\\
        0& 1 & 0\\ 
         -\mathrm{cos}(\theta)&0 & \mathrm{sin}(\theta)
    \end{pmatrix} 
\end{equation}

As a final step, the sky-plane-aligned model is projected onto the observed frame of the plane of the sky by integrating the three-dimensional emission model along the LoS. In our case, we employ a composite trapezoid rule algorithm \citep{carnahan_1969}. The LoS extent is arbitrarily tuned to span 100 elements on each side of the analysed galaxy, with an effective size 
of each pixel equal to the angular diameter extend of the pixels of each image. Given the limited extent of each galaxy in the sample with respect to the respective field of view (Sect.~\ref{sec:data_prep}) and, thus, of the expected physical depth, such a choice ensures that no appreciable emission is truncated out of the integration volume. Tests using different overall extents or discretization steps highlighted only negligible differences in the resulting projected maps, at the cost, though, of considerable increases in computing time.

\subsection{Fitting algorithm}
\label{sec:ins}
To perform model inference on the selected targets, we integrate the 3D thick disk model described in the previous section within a Bayesian forward modelling framework. In particular, we consider the implementation provided by the ``Source Characterization using a Composable Analysis''\footnote{\url{https://github.com/lucadimascolo/socca}} (\texttt{socca}; \citealt{socca}) library. A comprehensive description of the code base, of its optimization and functionalities, as well as of the forward model generation and posterior sampling can be found in the online documentation.\footnote{\url{https://lucadimascolo.github.io/socca}} We provide here a brief summary of the aspects of \texttt{socca} that are mostly relevant to this work.

In \texttt{socca}, source models are constructed by treating parametric surface brightness components as additive (and, optionally, interlinked) terms, allowing for a flexible description of sources with a complex morphology:
\begin{equation}
I(x,y) = \textstyle \sum_{k} I_k(x,y) .
\end{equation}
Here, each of the $I_k(x,y)$ terms refers to the specific surface brightness distribution of the $k$-th morphological component contributing to the overall source model. Following from the discussion in section~\ref{sec:methodology}, for the thick disk, $I_k = \int \textbf{R}_2 \textbf{R}_1 B(R,Z) \mathrm{d}z$, where $z$ denotes the LoS direction. The image resulting from the linear combination of the thick disk model with any additional feature (e.g., the bulge-like structure introduced further on in this work; see section~\ref{sec:fitting_real}) is then convolved with the instrumental PSF $P(x,y)$ to produce the predicted observation $\tilde{I}(x,y) = [P * I](x,y)$. For the sake of computational efficiency, we perform this via pointwise product of the Fourier-transformed model and PSF images. Given the moderate extent of the target sources and of the PSF with respect to the selected field of views (Sect.~\ref{sec:data_prep}), the non-periodicity of the maps introduces negligible effects on the resulting convolved images, making zero-padding or apodization unnecessary.

The prior probability distributions for each model parameter are specified by the user and sampled via a transformation from the unit hypercube to the prior phase space, computed through their respective inverse cumulative distribution function. For this purpose, \texttt{socca} relies on the implementation of standard probability distributions provided by the \texttt{numpyro.distributions} module \citep{numpyro_1, numpyro_2}. Along with conventional prior probabilities, \texttt{socca} supports tying parameters across model components --- e.g., enforcing shared centroid coordinates between concentric components --- and defining functional priors, in which a parameter is expressed as a user-specified function of other model parameters. This allows complex inter-parameter dependencies to be imposed directly at the prior level (e.g., when requiring the effective radius of one component to be a fraction of another; see the disk+bulge case in section~\ref{sec:fitting_real} below).

Given a sampled parameter set, the forward model generates a synthetic image to be compared with the observed data. We assume the noise in each pixel to be statistically independent and described by a multi-variate Gaussian distribution with diagonal covariance matrix, yielding the log-likelihood
\begin{equation}
    \textstyle\log\mathcal{L} = -\frac{1}{2} \sum_i [w_i (D_i-\tilde{I}_i)^2 + \log{2\pi/ w_i}]
\end{equation}
where the index $i$ runs over all the pixels of the model $\tilde{I}(x,y)$ and data $D(x,y)$. The inverse variance of the $i$-th pixel $w_i$ is computed from the local RMS estimates discussed in section~\ref{sec:RMS} for each galaxy cutout.

The model inference is performed using a nested sampling algorithm \citep{skilling_2004}, specifically employing the implementation provided by the \texttt{python} library \texttt{NAUTILUS}\footnote{\url{https://github.com/johannesulf/nautilus}} \citep{nautilus}. More broadly, the entire forward-modelling pipeline is built on \texttt{JAX} \citep{jax}, taking advantage of just-in-time compilation to accelerate the inference process.

\subsection{Mock data}
\label{sec:mockdata_gen}
To test our fitting code and define which data quality (i.e., angular resolution and S/N) is needed to accurately retrieve the scale height of the galaxies, we create mock data over a range of parameters that is representative of the real data.  

We generate mock data using equation (\ref{eq:brightnessprofile}) for galaxy masses in the range $3\times 10^9~\mathrm{M_{\odot}} - 10^{11}~\mathrm{M_{\odot}}$
and at 4 different redshifts: $z=\{1, 1.5, 2, 2.5\}$. To assign scale lengths to these galaxies, we assume the size-mass relation by \citet{2014vanderwel}. We then assign scale height to our mock galaxies, considering two categories: a thick disc and a thin disc. We take ratios of $(h_r/z_0)_{\mathrm{thick}}=2$, based on the the Milky Way thick disc \citep{Bland_Hawthorn_2016} and $(h_r/z_0)_{\mathrm{thin}}=9.4$ based on typical values observed in local thin discs \citep{2006_yoachim}. These ratios are chosen because they provide a large range of $z_0$ values, which will allow us to evaluate the lower limit of angular size that can be resolved. For the geometry of the galaxies, we assume inclination angles in the range $i=80^\circ - 90^\circ$ and position angles in the range $\theta=0^\circ - 125^\circ$ (see Appendix \ref{appedix:mockdata}).
Dependent on the galaxy mass and the redshift, some galaxies have a scale height that is smaller than the pixel scale of the image. These data are intentionally kept since we want to test the limit of the parameter recovery. We expect that any galaxy that is not resolved ($z_0\leq 1~\mathrm{pixel}$), will not be able to be fitted reliably. 

We then convolve each image with the \textit{JWST} PSF, generated using \texttt{WebPSF} \citep{perrin_2014}, corresponding to the filter that covers the rest-frame wavelength of 1.2 $\mu$m. We subsequently add noise such that, in the end, we have 160 galaxies in total (see Appendix \ref{appedix:mockdata}) with three distinct S/N categories ($\sim8$, $\sim12$, and $\sim25$) which we refer to as the low-, medium-, and high-S/N levels.

\subsection{Tests on the mock data}
\label{section:dataval}
We fit the mock data using the model described in Section~\ref{sec:3d}, which is defined by seven parameters: the galaxy centre ($x_\mathrm{c}, y_\mathrm{c}$), the radial scale length ($h_r$), the vertical scale height ($z_0$), the central intensity ($I_0$), the position angle ($\theta$), and the inclination ($i$).
To define the prior for $z_0$, we introduce a scaling factor $z_\text{scale}$ such that $z_0 =  h_r  z_\text{scale}$, where $z_\text{scale}$ is restricted to values between 0.05 and 1. This constraint ensures that the scale height remains smaller (or equal) than the radial scale length, thereby avoiding degeneracies between these two parameters. For the remaining parameters, we adopt uniform or log-uniform priors as appropriate. Table~\ref{tab:disc_prior} summarises the assumed prior ranges for all parameters.

\begin{table}
    \centering
    \begin{tabular}{ccc}
        \hline
        Parameter & Type of Prior & Range  \\ \hline
        $x_{\mathrm{c}}$  & uniform & RA' + $[-0.6, 0.6]$ (arsec) \\
        $y_{\mathrm{c}}$  & uniform & dec' +$[-0.6, 0.6]$ (arsec) \\
        $I_0$ & log uniform & $0.1:100$ (10.0 $\cdot$ nJy)\\
        $\theta$ & uniform &  $[-90^\circ,90^\circ]$  \\
        $i$& uniform  & $[72^\circ,90^\circ]$ \\
        $h_r$ &  uniform & [$0.1, 30]$ (kpc) \\
        $z_{\mathrm{scale}}$ & log uniform & $[0.05,1]$ \\
        $z_{0}$ & -  & $h_r\,z_{\mathrm{scale}}$  (kpc) \\\hline

    \end{tabular}
    \caption{Assumption on the prior type and range for the parameters defining the model.}
    \label{tab:disc_prior}
\end{table}

\begin{figure*}
    \centering
    \includegraphics[width=\linewidth]{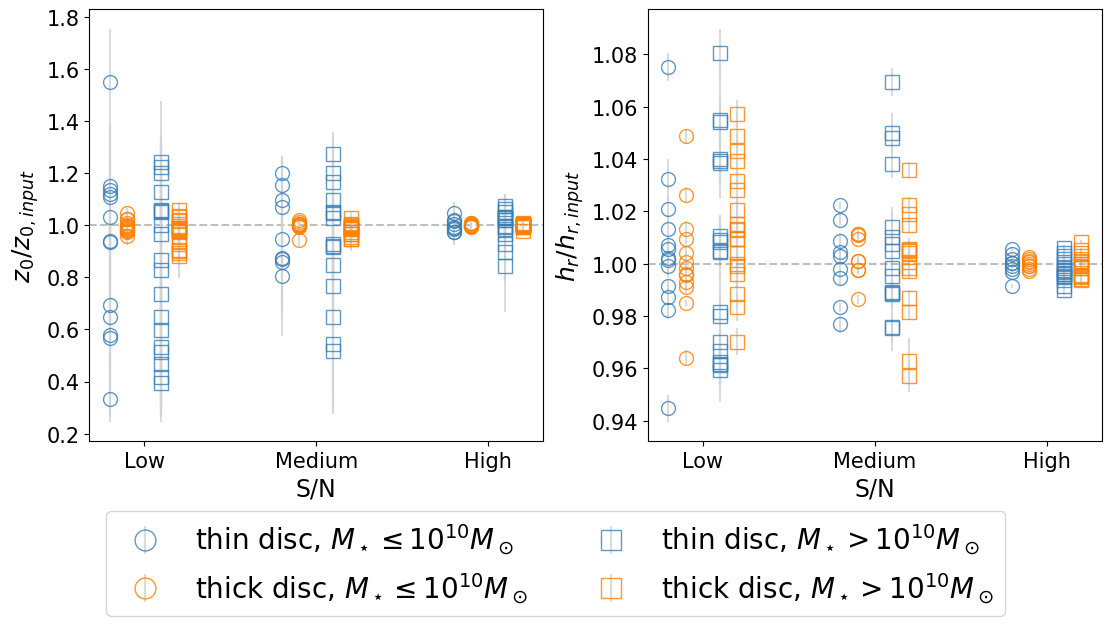}
    \caption{Recovery of the scale-height  ($z_0/ z_{0,\mathrm{input}}$), and scale length ($h_r/h_{r,\mathrm{input}}$) for mock data with different S/N ratios. Values of $z_0/ z_{0,\mathrm{input}}$ and $h_r/h_{r, \mathrm{input}}$ close to 1 (dashed horizontal lines) indicated perfect accuracy. The blue and orange markers show thin and thick discs, respectively. Low mass galaxies ($M_\star \leq 10^{10}~\mathrm{M_{\odot}}$) are shown with circles and high mass galaxies ($M_\star \geq 10^{10}~\mathrm{M_{\odot}}$) with squares.}
    \label{fig:full_sn.png}
\end{figure*}

In Fig.~\ref{fig:full_sn.png} we present the results of our tests, showing how well our tool recovers the scale length (left panel) and scale height (right panel) as a function of S/N. The ratio between the best-fit scale length and the input (“true”) value used to generate the mocks remains close to unity (see details in Appendix \ref{appedix:mockdata}), indicating that the radial scale length is robustly recovered over the full S/N range. We find similarly good performance for the inclination, as shown in Appendix \ref{appedix:mockdata}.

\begin{figure*}
    \centering
    \includegraphics[width=\linewidth]{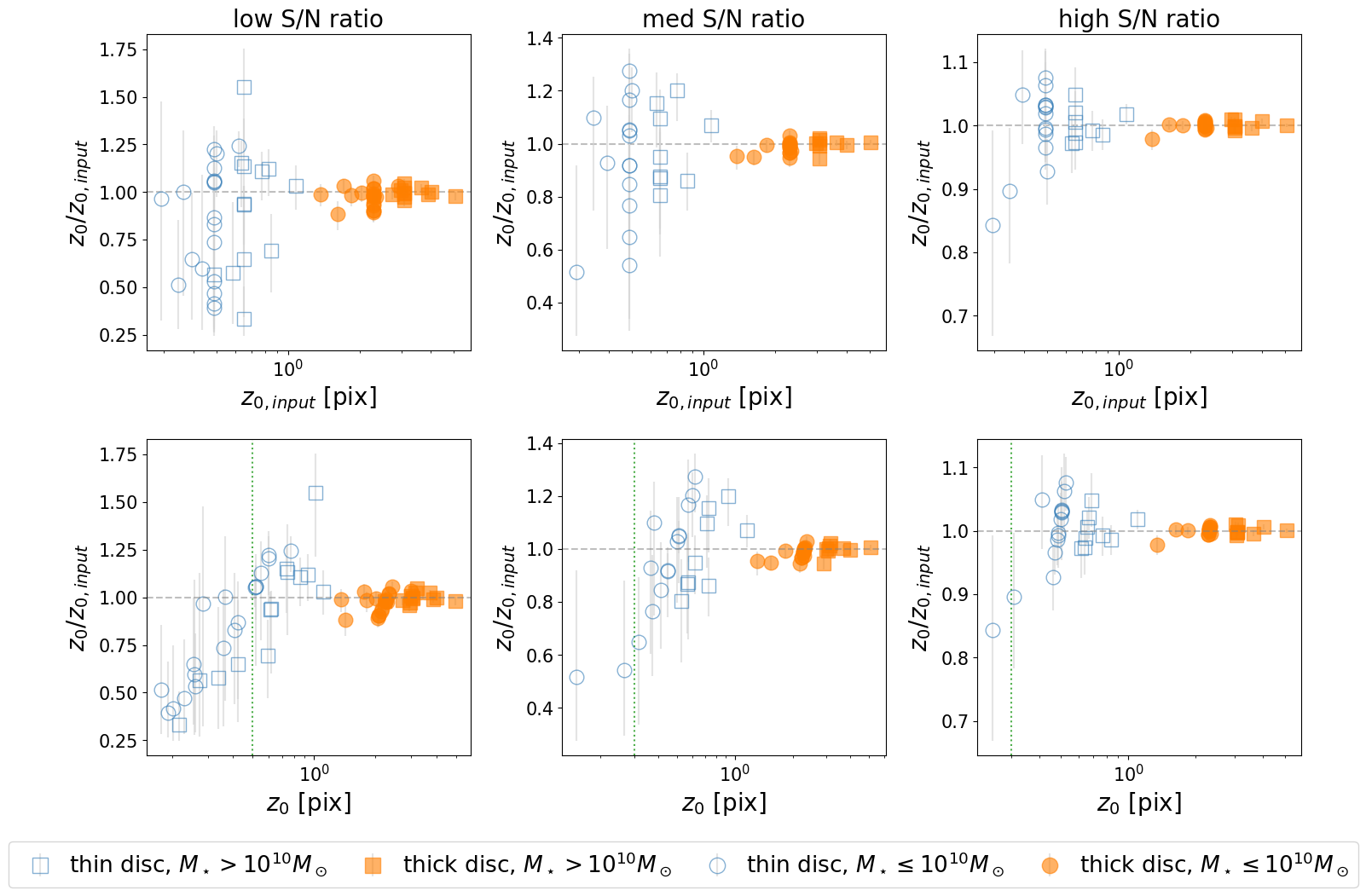}
    \caption{Recovery of the scale height  $z_{0}/ z_{0,\mathrm{input}}$ compared to the input (upper panels) and recovered (bottom panels) values of the scale height in pixel units for the sample of mock data. 
    The different columns show the results for the three different S/N levels. The circles correspond to low mass galaxies and the square to high mass galaxies for thin and thick discs, as indicated in the legend. The green dashed vertical lines indicate the locations of the thresholds below which we consider the fits as unreliable.}
    \label{fig:z_vs_pix}
\end{figure*}

The recovery of the vertical scale height is similarly stable for medium- and high-S/N galaxies: the ratio $z_{0}/z_{0,\mathrm{input}}$ typically lies in the range $0.6$–$1.5$, consistent with a reliable reconstruction of the disc thickness. The performance degrades for low-S/N, low-mass ($\leq 10^{10}\,\mathrm{M_{\odot}}$) thin discs, where the model can underestimate %overestimate 
$z_0$ by up to a factor of $\sim$2.
To understand why the thickness becomes less reliable in this regime, and to disentangle the effects of resolution and S/N, Fig.~\ref{fig:z_vs_pix} shows the recovery of $z_0$ as a function of the recovered value $z_0$ (upper panels) and the input value $z_{0,\mathrm{input}}$ (lower panels). For most mock galaxies, $z_{0}/z_{0,\mathrm{input}} \approx 1$, indicating that the intrinsic parameters are recovered accurately in the majority of cases. A clear resolution-driven limit appears at $z_{0} \sim 0.3$ pixels (equivalently $z_{0,\mathrm{input}} \sim 1$ pixel), below which $z_0$ is biased low by roughly a factor of two. For low-S/N galaxies, this threshold shifts to slightly larger values, around $z_{0} \sim 0.5$ pixels. We therefore adopt recovery limits of $z_{0} \geq 0.3$ pixels for medium- and high-S/N images, and $z_{0} \geq 0.5$ pixels for low-S/N cases.

To summarise, our mock galaxy analysis shows that the angular resolution limit for reliably recovering the structural parameters depends on the image S/N, with the scale height setting the effective limiting angular size. Based on these results, we adopt thresholds of $z_0 \geq 0.5$~pixels for low-S/N data (S/N~$<10$) and $z_0 \geq 0.3$~pixels for medium- and high-S/N data (S/N~$>10$).


\subsection{Fitting the sample}
\label{sec:fitting_real}
In Section \ref{section:dataval} we described the fitting procedure for the mock data and defined the prior distributions for all the model parameters. As the mock galaxies are generated with values consistent with observations, we use the same priors for fitting the disc of real galaxies (see Table \ref{tab:disc_prior}).

In addition to the disc-only fit based on the surface-brightness profile in equation~(\ref{eq:brightnessprofile}), we also fit each galaxy with an extra Sérsic component, equation~(\ref{eq:sersic}), to model a central bulge. For each object, we select the preferred model by evaluating the residuals and through visual inspection. In fact, although Bayesian evidence estimates are directly available as a byproduct of the nested sampling algorithm, the significance of the residuals alone provide sufficiently clear indications of whether an additional bulge component is necessary, making a formal Bayesian model comparison redundant. Including a bulge is required for galaxies with a prominent central excess that cannot be reproduced by a single-component disc profile (see Section~\ref{sec:results}). 

For the bulge component, we assume that the centre and position angle parameters are the same of the disc. We also impose a prior on the bulge effective radius, $R_\mathrm{e}$, expressed in terms of the disc scale length $h_r$:
$R_\mathrm{e} = R_{\mathrm{scale}} \,1.678 \, h_r$,
where $R_{\mathrm{scale}}$ is a dimensionless scaling factor constrained to $0 \le R_{\mathrm{scale}} \le 1$. The factor 1.678 converts an exponential scale length to an effective radius. This prior ensures that the Sérsic component remains confined to the inner regions and therefore captures the bulge rather than the outer disc.

After fitting the galaxies in our sample, we find that 36 systems have best-fit scale heights $z_0 < 0.3$ pixels. Our mock tests indicate that robust recovery requires $z_0 \ge 0.5$ pixels for low--S/N images and $z_0 \ge 0.3$ pixels for medium- to high--S/N images. Of these 36 galaxies, 34 have medium to high S/N and 2 have low S/N. We therefore refit the 34 medium- to high--S/N systems while fixing $z_0 = 0.3$ pixels, and refit the two low--S/N systems while fixing $z_0 = 0.5$ pixels. The mock analysis further shows that when the best-fit value falls below the relevant threshold, $z_0$ is typically underestimated by up to a factor of two. We thus treat these measurements as upper limits and adopt $z_0 = 2 \times 0.3$ pixels and $z_0 = 2 \times 0.5$ pixels for the medium- to high--S/N and low--S/N cases, respectively. In addition, for 19 galaxies, we find $z_{\mathrm{scale}}$ values at the lower edge of the prior, with corresponding $z_0$ values close to 0.3 pixels. We refit these galaxies using the same approach, fixing $z_0$ to 0.3 pixels. As before, the resulting $z_0$ values are treated as upper limits.

\section{Results}
\label{sec:results}
\begin{figure*}
    \centering
    \includegraphics[width=\linewidth]{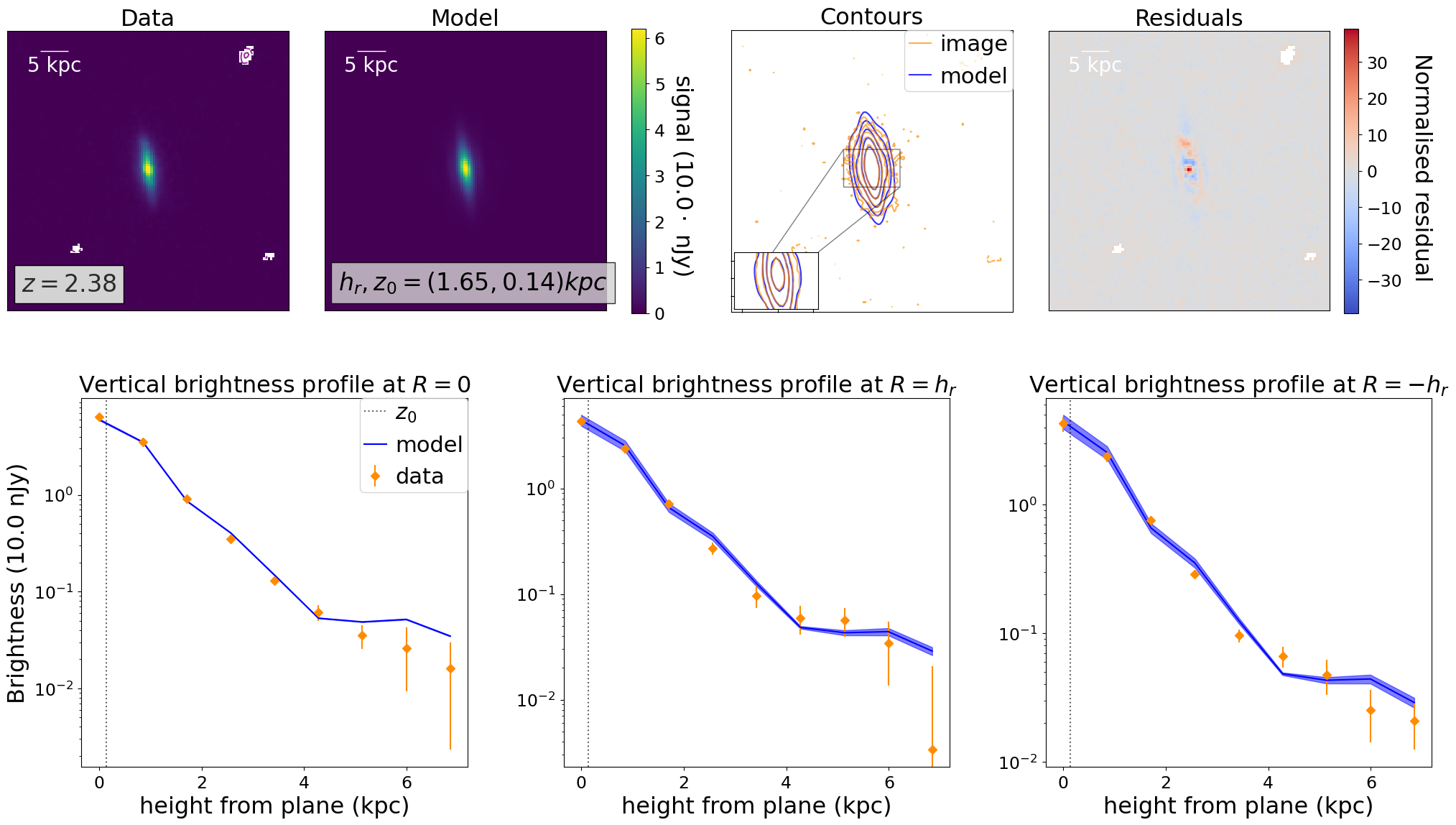}
    \caption{Representative example of a galaxy (ID26) that is well fitted by a single disc component. Upper panels, from left to right: masked data with contours of the masked regions in purple; best-fit model; contours of the data (blue) and model (orange); residuals normalised to the RMS. Bottom panels, from left to right: vertical brightness profile of the data (orange points) and the model (shaded curve) at $R = 0$, $R = h_r$, and $R = -h_r$. }
    \label{fig:fit-disc}
\end{figure*}

\begin{figure*}
    \centering
    \includegraphics[width=\linewidth]{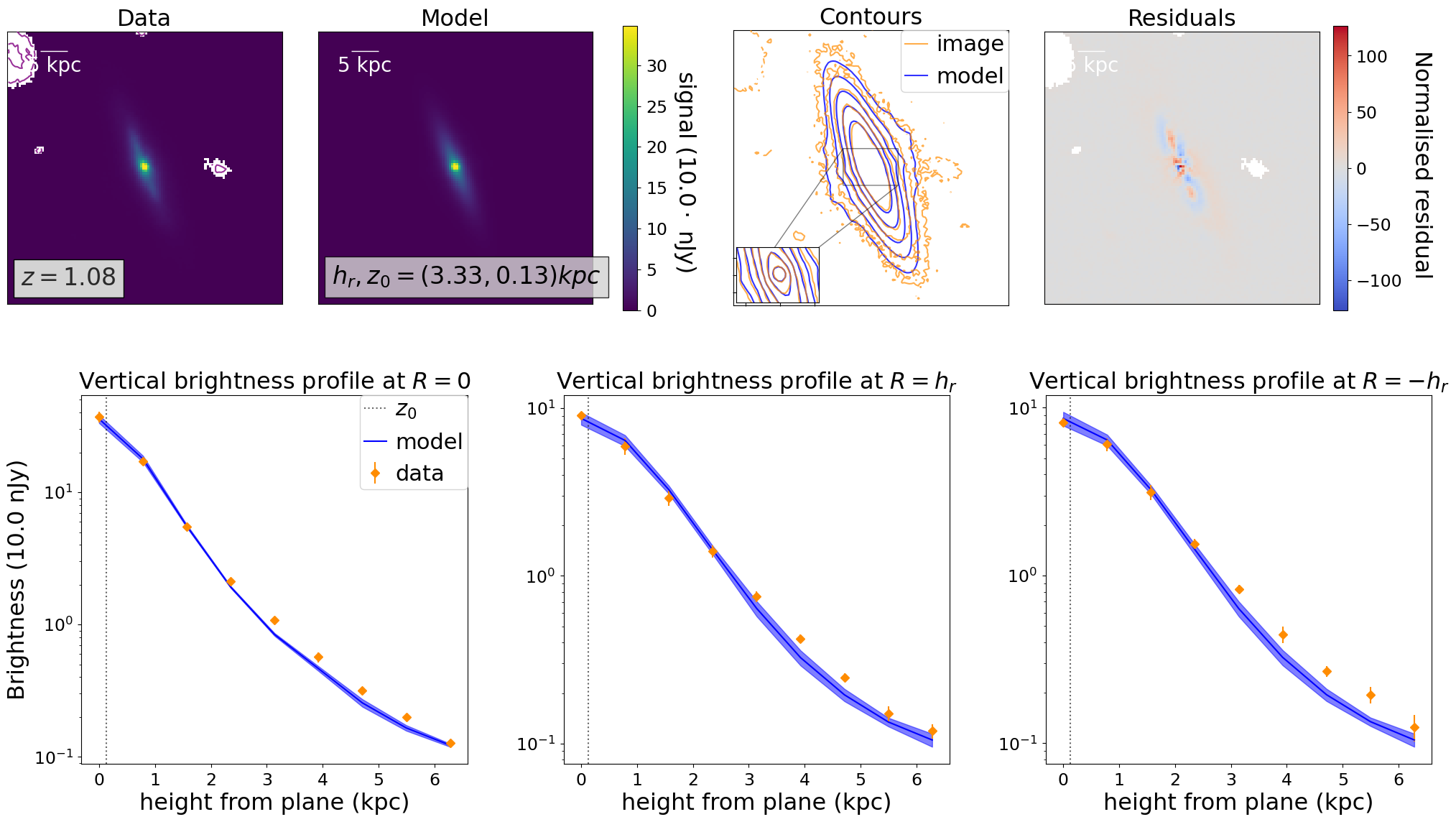}
    \caption{Same as Fig. \ref{fig:fit-disc} but for a galaxy (ID17) that is well fitted with a disc and a bulge component.}
    \label{fig:fit-bulge}
\end{figure*}

Of the 90 galaxies in the sample, 46 are best fitted by a single-disc model, 44 are best fitted by a disc + bulge model. Fig. \ref{fig:fit-disc} shows an example of a galaxy best fitted by a single disc, and Fig. \ref{fig:fit-bulge} illustrates a case where a disc + bulge model provides a better fit. In both figures, the top row shows the data and corresponding model, together with their contours and residuals. The residuals are obtained by subtracting the model from the data and normalizing by the noise. The first panel shows the original image along with the masked regions.
The bottom row presents the vertical brightness profiles of both the data and the model. The brightness profile is extracted perpendicular to the galaxy’s radial axis. For each pixel $j$, the signal is evaluated at a height $Z(j)$ above the galactic plane. The profile is measured using a strip of three pixels centred at radius $R$, with the signal estimated as the mean and its uncertainty given by the standard deviation. The first panel shows the brightness profile at the galaxy centre ($R=0$), while the central and right panels correspond to profiles at $R=+h_r$ and $R=-h_r$. Analogous figures for all galaxies are provided in the Supplementary material. A table of best-fit parameters for the full sample will be made available via VizieR.

Among the 90 galaxies, 55  (27 fitted with a disc model and 28 with a disc + bulge model) yield values of $z_{0}$ below the lower limit of $z_0 \geq 0.3$ pixel, or at the lower edge of the prior for $z_0/h_r$. These galaxies were refitted with a fixed values for $z_0$ as described in Section \ref{sec:fitting_real}, and will be considered as upper limits.


\begin{figure*}
    \centering
    \includegraphics[width=\linewidth]{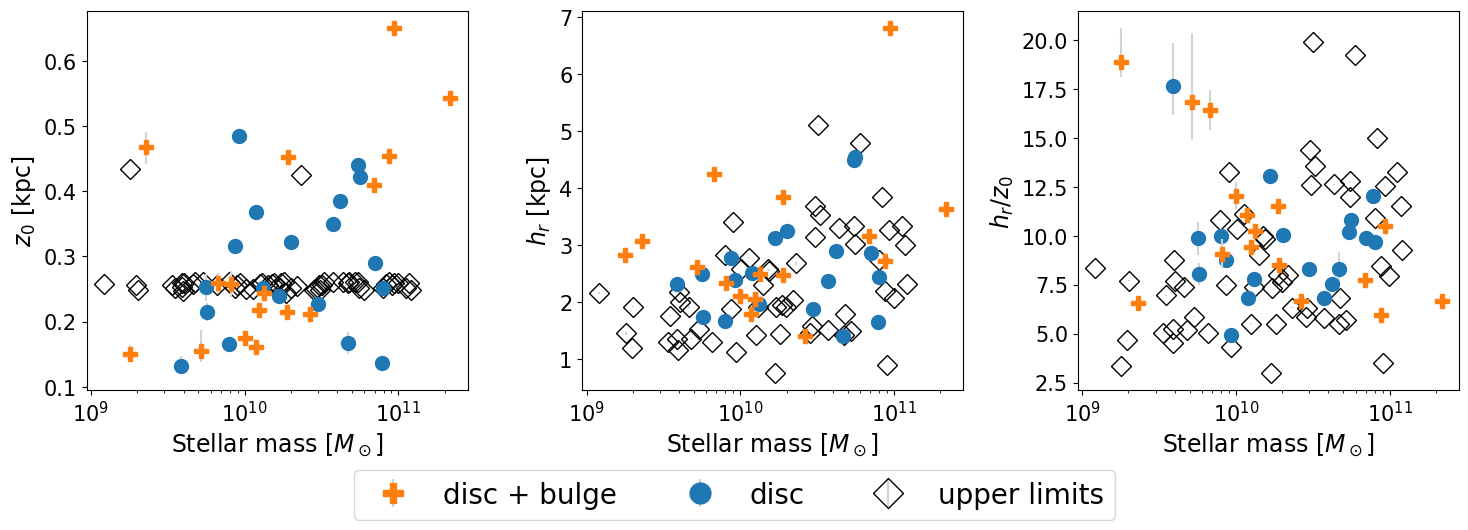}
    \caption{Best-fit values of $z_0$, $h_r$ and the ratio $h_r$/$z_0$  as a function of their stellar masses $M_{\star}$. The orange crosses show galaxies that are fitted with a disc + bulge model, the blue circles show galaxies fitted with a single disc. The diamond markers show galaxies for which we have only an upper limit on $z_0$. }
    \label{fig:results-mass}
\end{figure*}

Fig. \ref{fig:results-mass} presents the best-fit values of $z_0$, $h_r$, and their ratio as a function of stellar mass for all galaxies in our sample. Galaxies best fitted by a single-disc model are shown as blue circles, while those fitted with a disc + bulge model are represented by orange crosses. Empty symbols indicate galaxies for which only an upper limit on $z_0$ is available.
The distribution of galaxies in the $z_0$ and $h_r/z_0$ versus $M_\star$ planes shows no clear trends between these quantities and stellar mass. The scatter in both $z_0$ and $h_r/z_0$ appears roughly uniform across the full mass range. In contrast, $h_r$ exhibits a mild increase with stellar mass, consistent with the expected mass–size relation \citep[e.g.][]{2014vanderwel}.

\begin{figure*}
    \centering
    \includegraphics[width=\linewidth]{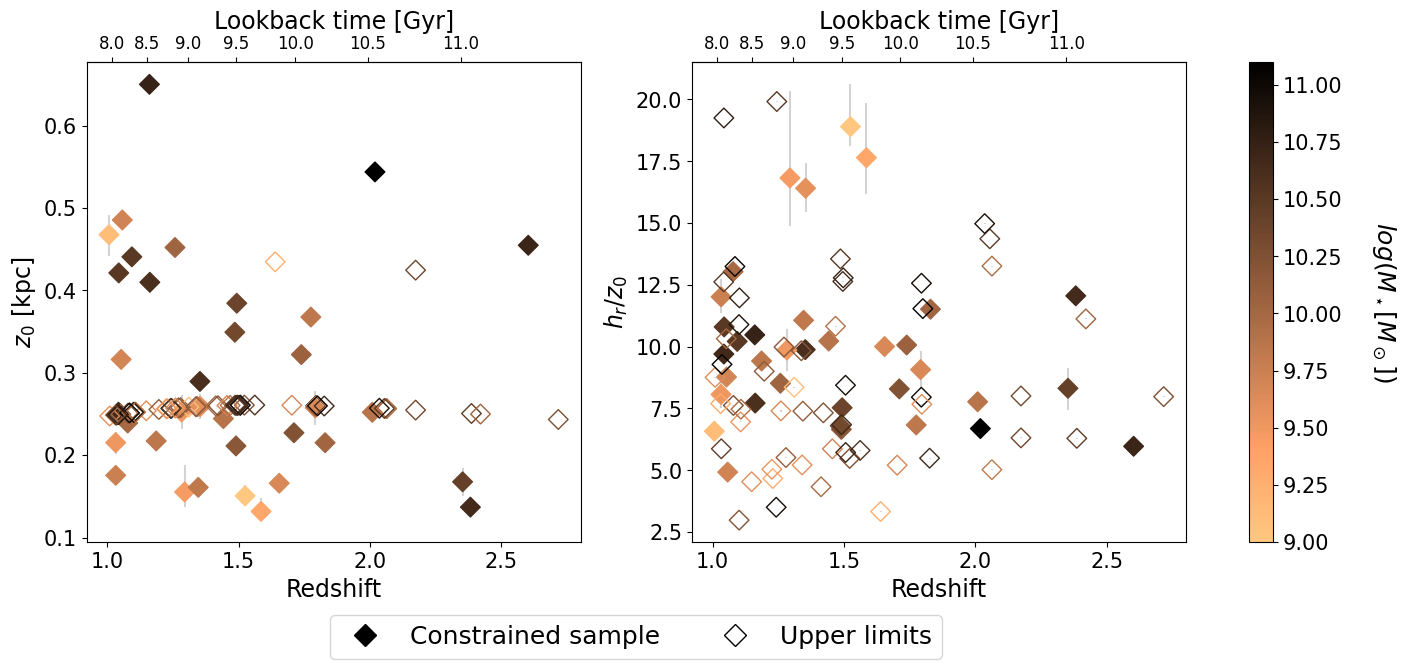}
    \caption{Best-fit values of $z_0$, $h_r$ and the ratio $h_r$/$z_0$  as a function of redshift. The markers are colour coded according to $M_\star$, as indicated by the colour bar. The empty markers are the fits where we only have an upper limit for $z_0$. The top axis shows the associated look-back time of the galaxies.}
    \label{fig:vs_z_mass}
\end{figure*}

Next, we examine possible trends with redshift. Fig.~\ref{fig:vs_z_mass} shows $z_0$ and $h_r/z_0$ as a function of redshift, with points coloured by stellar mass $M_\star$. None of the parameters show evidence of a correlation with stellar mass across redshift. While small values of $z_0$ are found at all redshifts, the upper envelope of $z_0$ decreases toward higher redshift, such that the largest scale heights observed at lower redshift are not present at higher redshift. In contrast, the $h_r/z_0$ ratios show a similar spread at all redshifts, with no clear systematic evolution. This suggests that $z_0$ and $h_r$ evolve proportionally with cosmic time.

\begin{figure*}
    \centering
    \includegraphics[width=\linewidth]{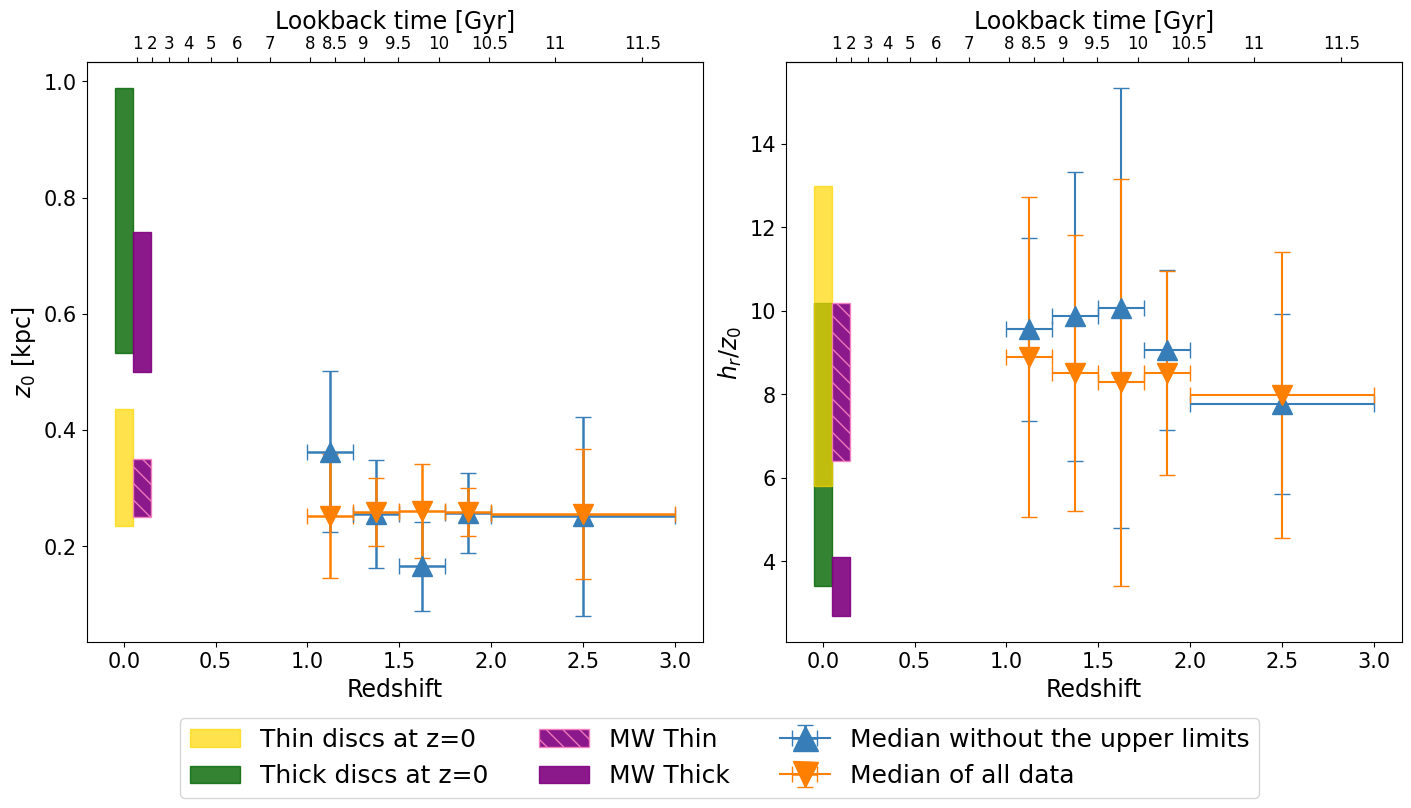}
    \caption{Median values of $z_0$ and $h_r$/$z_0$ as a function of redshift when the galaxies for which we have upper limits on $z_0$ are excluded (blue upper triangles) or not (orange lower triangles). The bars in the x-direction show the extend of the redshift bin and the bar in the y-direction show the standard deviations of the distribution. The yellow and green rectangular shapes show the values of $z_0$ and $h_r$/$z_0$ for the thick and thin discs at $z = 0$ \citep{2006_yoachim}, while the purple and hatched regions show the corresponding values for the Milky Way \citep{Cheng_2012,Bland_Hawthorn_2016,2018_mateu,Tkachenko_2025}}.
    \label{fig:vs_z}
\end{figure*}

To investigate potential redshift evolution in greater detail, we compute the median values of $z_0$ and $h_r/z_0$ within redshift bins, as shown in Fig. \ref{fig:vs_z}. We consider two cases: the median excluding upper-limit values (blue triangles) and the median including all data points (orange triangles). Horizontal error bars represent the width of the redshift bins, while vertical bars indicate the $1\sigma$ scatter, computed considering the standard deviation of the distributions. Redshift bins are spaced by $\Delta z = 0.25$ up to $z = 2$; beyond this, the bin width is increased to $\Delta z = 1$ due to the limited number of galaxies at higher redshift. The highest-redshift bin contains only massive systems ($M_\star > 10^{10}~\mathrm{M_\odot}$). The corresponding median values of $z_0$ and $h_r/z_0$, together with their intrinsic scatter, are listed in Table \ref{tab:z_medians}. In Fig. \ref{fig:vs_z}, we also show the range of $z_0$ and $h_r/z_0$ values derived from local studies \citep{2006_yoachim} and for the Milky Way \citep{Cheng_2012, Bland_Hawthorn_2016, 2018_mateu, Tkachenko_2025}, for both the thin and thick discs. \citet{2006_yoachim} analysed the structural parameters of 36 nearby edge-on spiral galaxies in the B, R, and K$_S$ bands, fitting each with both single- and double-disc models using a $\mathrm{sech}^2$ vertical profile. For comparing our data with thick and thin discs in the local universe, we adopt their double-disc parameters measured in the R band, correcting for the difference between $\mathrm{sech}^2(Z/z_0)$ and $\mathrm{sech}^2(Z/2z_0)$ by dividing their scale heights by a factor of two. 
For the Milky Way thin disc, we assume a scale height of $z_{0,\mathrm{thin}} = 0.3 \pm 0.05~\mathrm{kpc}$ and a scale length of $h_{\mathrm{r,thin}} = 2.5 \pm 0.4~\mathrm{kpc}$ \citep{Bland_Hawthorn_2016}. For the thick disc, we adopt $z_{0,\mathrm{thick}} = 0.62 \pm 0.12~\mathrm{kpc}$ and $h_{\mathrm{r,thick}} = 2.10 \pm 0.18~\mathrm{kpc}$, based on recent studies \citep{Cheng_2012, Bland_Hawthorn_2016, 2018_mateu, Tkachenko_2025}. 

Across all redshift bins, the median values of both $z_0$ and $h_r/z_0$ in our sample are consistent, within the scatter, with those of thin discs in the local Universe, including the Milky Way. In the local sample of \citet{2006_yoachim}, the $h_r/z_0$ ratios of thin and thick discs overlap substantially, limiting the discriminating power of this ratio alone. Nevertheless, our results show the same overall behaviour whether the medians are computed including upper limits (orange downward triangles) or excluding them (blue upward triangles). Overall, this comparison suggests that the majority of galaxies in our sample are structurally consistent with local thin discs across the full redshift range probed.

\begin{table*}
    \centering
     %\resizebox{\textwidth}{!}
     \begin{tabular}{|c|ccc|ccc|}
        \hline
         & \multicolumn{3}{|c|}{All Data Points} & \multicolumn{3}{|c|}{Excluding upper limits}\\
        \hline
        redshift  & N & $z_{\mathrm{0}}$ [kpc] & $h_r/z_0$ & N  & $z_{0}$ [kpc]  & $h_r/z_0$ \\ \hline
        $1.0-1.25$&32&$0.25 \pm0.11$&$8.9\pm3.8$&12&$0.36\pm0.14$&$9.6\pm 2.2$\\
        $1.25-1.5$&25&$0.26 \pm0.06$&$8.5\pm3.3$&10&$0.26\pm0.09$&$9.9\pm 3.5$\\
        $1.5-1.75$&11&$0.26 \pm0.08$&$8.3\pm4.9$&5&$0.17\pm0.08$&$10.1\pm 5.3$\\
        $1.75-2.0$&8&$0.26 \pm0.04$&$8.5\pm2.4$&3&$0.26\pm0.07$&$9.1\pm 1.9$\\
        $2.0-3.0$&14&$0.26 \pm0.11$&$8.0\pm3.4$&5&$0.25\pm0.17$&$7.8\pm 2.1$\\
        $1-3$&90&$0.26 \pm0.09$&$8.4\pm3.7$&35&$0.25\pm0.14$&$9.7\pm 3.3$\\

        \hline
    \end{tabular}
    \caption{Number of galaxies, median values and corresponding standard deviations of $z_0$ and $h_r/z_0$ per redshift bin. The left columns show the median values when taking into account all the data points, while the right columns give the values for the subsmaple obtained after excluding galaxies for which we have only upper limits on $z_0$. }
    \label{tab:z_medians}
\end{table*}

\begin{figure*}
    \centering
    \includegraphics[width=\linewidth]{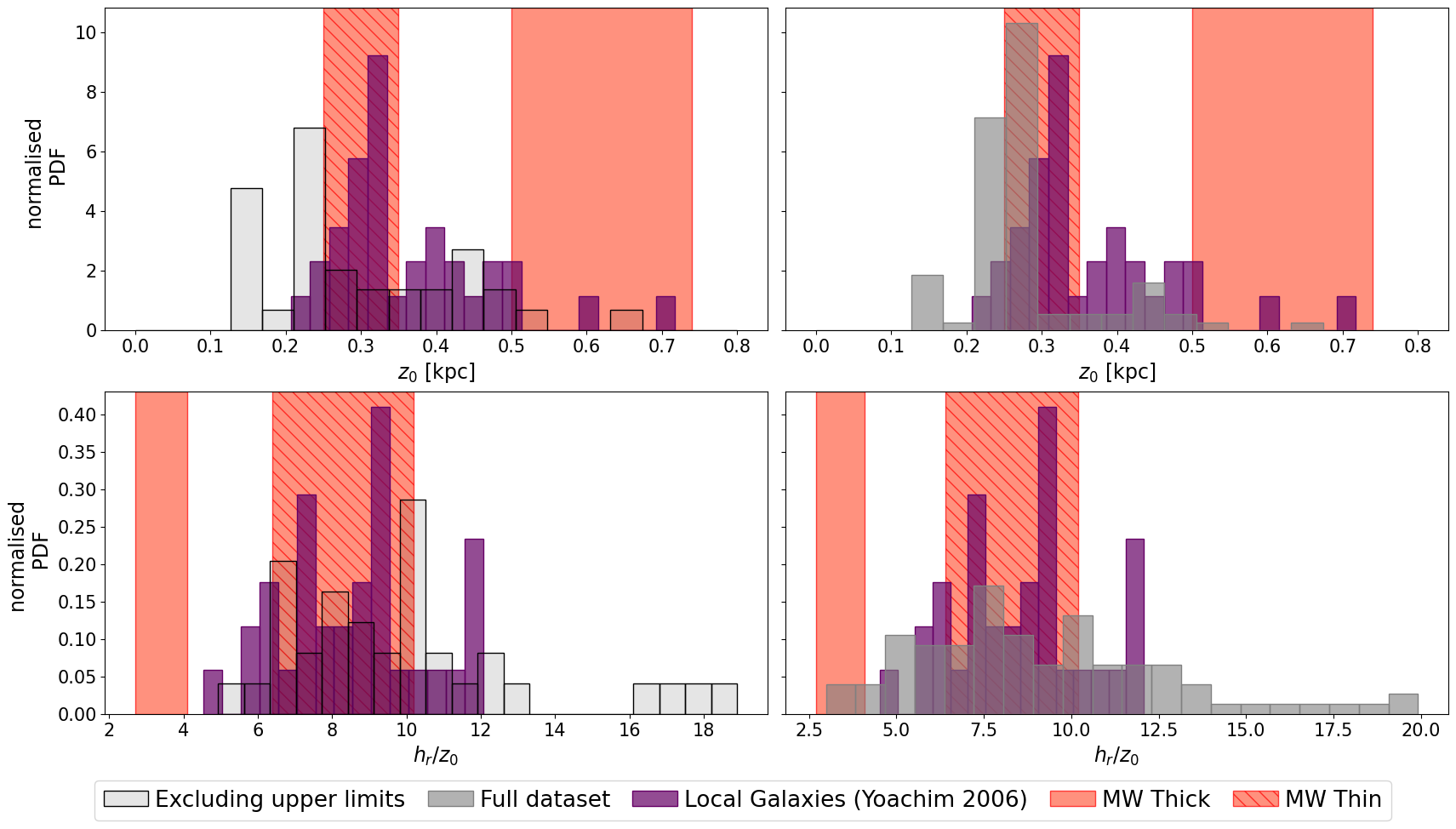}
    \caption{Normalised distributions of the scale height $z_0$ (upper panels) and the axis ratio $h_r/z_0$ (lower panels) for our sample compared to local galaxies. Left panels: distributions excluding upper limits. Right panels: full distributions including upper limits. Purple bars show $z=0$ galaxies \citep{2006_yoachim}; red hatched and solid bars show the Milky Way thin \citep{Bland_Hawthorn_2016, Xiang_2018, 2023viera} and thick disc \citep{Bland_Hawthorn_2016, Tkachenko_2025}, respectively.}
    \label{fig:local_hist}
\end{figure*}

Fig.~\ref{fig:local_hist} shows the normalised distributions of $z_0$ and $h_r/z_0$. The open black histograms show our best-fit parameters excluding upper limits, while the grey histograms include them. For comparison, the purple bins indicate single-disc measurements for local galaxies from \citet{2006_yoachim}. Because in local galaxies the thick disc typically contributes only a fraction of the total disc luminosity, single-disc fits tend to recover scale heights closer to those of the thin disc; accordingly, the ratio between single-disc and thin-disc $z_0$ has a median of $1.3_{-0.1}^{+0.3}$. The red shaded regions mark the ranges spanned by the Milky Way thin and thick discs.
Overall, the $z_0$ and $h_r/z_0$ distributions of our sample broadly resemble those of local galaxies, but they exhibit a pronounced tail toward smaller $z_0$ and larger $h_r/z_0$. Quantitatively, the local sample has median single-disc values of $z_0 = 0.41^{+0.13}_{-0.11}~\mathrm{kpc}$ and $h_r/z_0 = 8.6^{+1.6}_{-1.7}$, compared to $z_0 = 0.26 \pm 0.09~\mathrm{kpc}$ and $h_r/z_0 = 9.7 \pm 3.3$ for our sample, i.e. local discs are thicker by a factor of $\sim 1.6$ and have slightly smaller $h_r/z_0$ by a factor of $\sim 1.2$. The extended tail of our sample therefore indicates the presence of discs that are significantly thinner than typical local discs. Relative to the Milky Way, our distributions are broadly consistent with the thin disc, while the Milky Way thick disc overlaps only with the two largest $z_0$ values in our sample.

\section{Comparison with previous high-z studies}
\label{sec:comparison}
In this section, we compare our results with previous estimates of disc thickness obtained from \textit{HST} data and \textit{JWST} data at $z > 0$.

\begin{figure*}
    \centering
    \includegraphics[width=\linewidth]{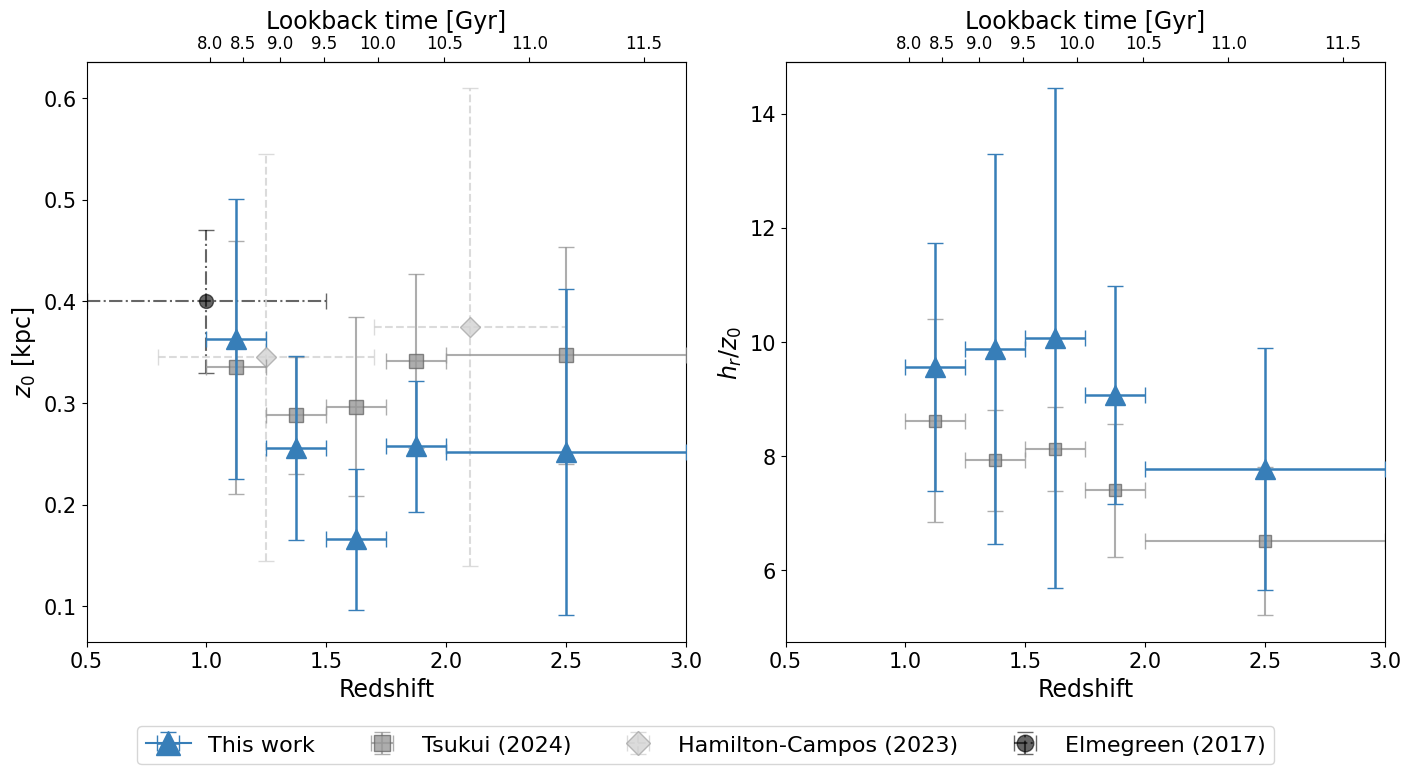}
    \caption{Median values of the scale height over redshift bins for our sample (blue triangles), compared to previous studies based on \textit{HST} \citep{Elmegreen_2017, 2023-Hamilton-campos} and \textit{JWST} \citep{2024Tsukui} data. }
    \label{fig:Comparison}
\end{figure*}

\citet{Elmegreen_2017} used \textit{HST} imaging in the F435W, F606W, and F814W filters, corresponding to a wavelength range of approximately $0.4$–$0.8~\mathrm{\mu m}$. Their sample includes 107 galaxies spanning redshifts $z = 0.5$–4, most with stellar masses between $10^{9.5}$ and $10^{10.5}~\mathrm{M_{\odot}}$. Within this sample, they distinguish between 35 spiral galaxies and 72 clumpy galaxies, with all spiral galaxies found at $z < 1.5$.
To model the vertical structure of these galaxies, \citet{Elmegreen_2017} extracted only the vertical surface brightness profiles, without fitting the radial component or disc scale length. They accounted for PSF convolution using the line spread function. The authors report\footnote{The value reported by \citet{Elmegreen_2017} is 0.8 kpc, which we divide by two to correct for the difference between the $\mathrm{sech}^2(Z/z_0)$ profile they adopted and the $\mathrm{sech}^2(Z/2z_0)$ definition used in this work.} a median disc thickness of $0.4 \pm 0.07~\mathrm{kpc}$ within the redshift range $z = 0.5$–1.5. They find that scale height increases with galaxy mass for both morphological types across all redshifts, but no significant correlation is found between $z_0$ and redshift.

\citet{2023-Hamilton-campos} analysed disc thicknesses over $z = 0.4$–2.5 using \textit{HST} imaging of the GOODS–South field at a fixed rest-frame wavelength of $0.46$–$0.66~\mathrm{\mu m}$. Their sample comprises 491 galaxies with stellar masses between $10^9$ and $10^{11}~\mathrm{M_{\odot}}$. They modelled the vertical profiles using a similar 1D $\mathrm{sech}^2$ function as \citet{Elmegreen_2017} and report\footnote{We divide their reported value of 0.71 kpc by two to account for the difference in the $\mathrm{sech}^2$ definition.} a median scale height of $z_0 = 0.36 \pm 0.18~\mathrm{kpc}$, finding no strong trend with redshift. They conclude that high-redshift galaxies are, on average, thinner than present-day discs, with the largest $z_0$ values occurring at lower redshifts. Their main conclusion is that thick discs form already thick and continue to thicken over time, with most of this evolution occurring at $z \lesssim 1$.

Fig.~\ref{fig:Comparison} compares the median $z_0$ values obtained in this work with those from previous studies. The blue markers represent the median of our sample (excluding upper limits) in redshift bins matched to those used by \citet{Elmegreen_2017} and \citet{2023-Hamilton-campos}. The light grey diamonds show the results from \citet{2023-Hamilton-campos}, and the black circles indicate the medians from \citet{Elmegreen_2017}.

Our measured scale heights are systematically lower than those reported in these earlier studies. Several factors may explain this discrepancy. First, \textit{JWST} offers superior spatial resolution compared to \textit{HST}: in comparable filters ($\sim 1.1~\mathrm{\mu m}$), \textit{JWST} has a PSF FWHM of 0.037 arcsec, while the \textit{HST} PSF has a FWHM of 0.089 arcsec. At $z = 1$, this corresponds to physical resolutions of approximately 0.3 and 0.7 kpc, respectively, a substantial difference given that measured $z_0$ values are of the same order. Second, the modelling approaches adopted by \citet{Elmegreen_2017} and \citet{2023-Hamilton-campos} (see also Introduction) likely bias their inferred thicknesses, as these are derived by fitting vertical profiles independently, without 
accounting for flux contributions from neighbouring pixels. Our methodology, which explicitly incorporates the PSF, is therefore expected to yield more reliable measurements. Finally, differences in rest-frame wavelength may also play a role. Our measurements are based on NIR imaging rather than optical, which traces older, and typically thicker, stellar populations. Despite this, we find systematically thinner discs, reinforcing the conclusion that earlier measurements likely overestimated $z_0$ due to resolution and modelling limitations.

\begin{figure*}
    \centering
    \includegraphics[width=\linewidth]{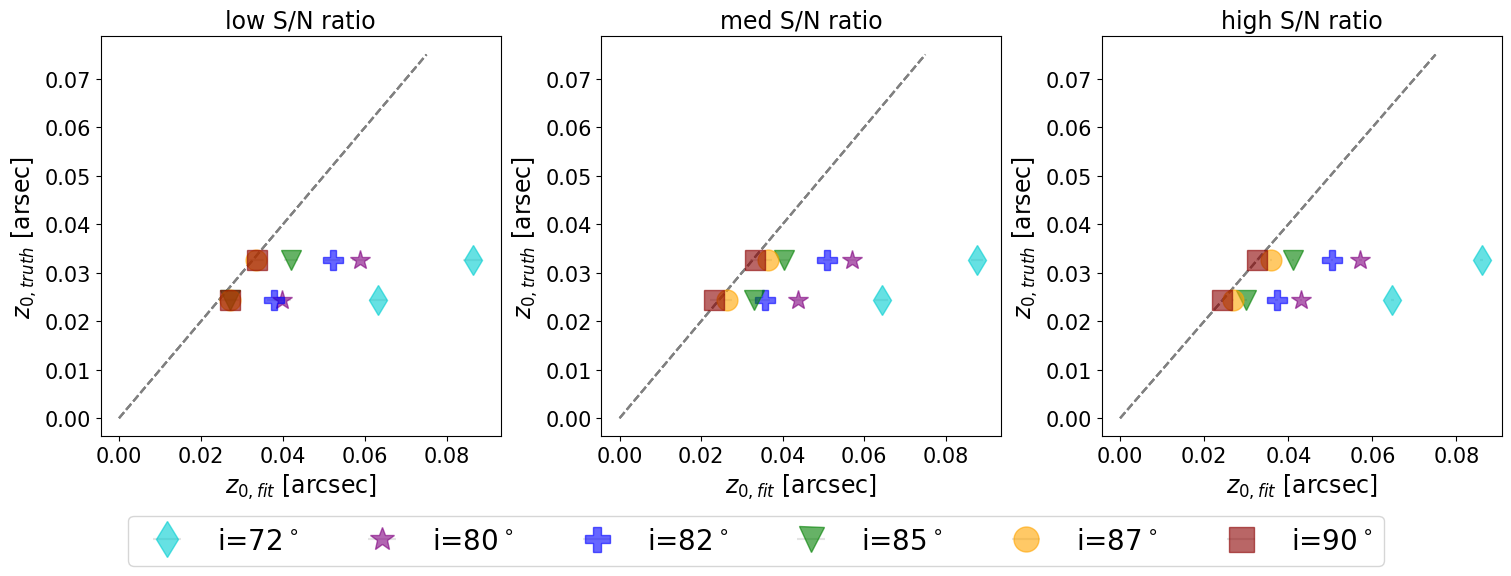}
    \caption{Intrinsic vs recovered scale-height for mock data with low-, medium and high S/N ratio, as obtained when galaxies with inclination above 80$^\circ$ are assumed perfectly edge-on (inclination of 90$^\circ$). The dashed gray line shows the 1:1 relation. }
    \label{fig:inclination_fixed}
\end{figure*}

As discussed in the Introduction, in recent years, two studies have investigated the thickness of high-redshift galaxies using \textit{JWST} data \citep{2024_lian,2024Tsukui}. \citet{2024_lian} employed a 1D approach to measure the vertical scale heights of galaxies at $z = 0 - 5$. They found that galaxies at $z > 1.5$ show no clear dependence of scale height on redshift, with median values in remarkable agreement with that of the Milky Way’s thick disc. In contrast, galaxies at $z < 1.5$ exhibit a systematic decrease in disc thickness toward lower redshift, with low-redshift galaxies having scale heights comparable to that of the Milky Way’s thin disc. Based on these results, they concluded that local thick discs were already thick at the time of their formation, and that secular heating is unlikely to be the dominant mechanism driving their evolution. However, \citet{2024_lian} did not provide median values for their sample nor individual thickness measurements. For this reason, we perform a quantitative comparison only with the study by \citet{2024Tsukui}, who adopted a fully 3D modelling approach including both thin and thick disc components.
\citet{2024Tsukui} used \textit{JWST} imaging to measure the vertical structure of 111 disc galaxies spanning redshifts $z = 0.1$–3, with stellar masses between $10^{8.5}$ and $\sim10^{11}~\mathrm{M_{\odot}}$. Their analysis employed rest-frame wavelengths of approximately $2.2~\mathrm{\mu m}$ for galaxies at $z < 1.45$, $1.6~\mathrm{\mu m}$ for $1.45 < z < 2.24$, and $1.2~\mathrm{\mu m}$ for $2.24 < z < 3$. They adopted the same brightness profile used in this work, equation~(\ref{eq:brightnessprofile}), and fitted each galaxy using one of four models: (1) a single disc, (2) a thin + thick disc, (3) a single disc + bulge, or (4) a thin + thick disc + bulge. The bulge component was modelled with a Sérsic profile.
They found that 67 galaxies were best fitted with single-disc models rather than double-disc configurations. Among these, 39 galaxies ($\sim58\%$) were well described by a single disc + bulge model, a fraction comparable to ours (46 out of 90, $\sim51\%$). \citet{2024Tsukui} observed a mild decrease in $z_0$ toward higher redshift, with no discernible trend in the ratio $h_r / z_0$. Their main conclusion is that thick discs form already thick and that thin discs are subsequently assembled at later stages.
Fig.~\ref{fig:Comparison} presents the median values of $z_0$ and the axis ratio $h_r / z_0$ as a function of redshift. The dark blue markers represent the medians from our sample (excluding upper limits), while the dark grey squares show the median values from \citet{2024Tsukui} over the same redshift bins. To enable a one-to-one comparison, we consider only the 67 galaxies in their sample that were best fit by single-disc (with or without bulge) models, excluding double-disc fits.
The left panel of Fig.~\ref{fig:Comparison} shows that for $z \leq 1.5$, our scale height measurements are consistent with those of \citet{2024Tsukui}. At higher redshifts, however, their values are systematically larger than ours by $\sim 30 \%$. In contrast, the right panel shows that our $h_r/z_0$ ratios are systematically higher across all redshift bins. These systematic differences likely arise from differences in the adopted modelling assumptions \citep{1997_degrijs}. A key distinction between our method and the one employed by \citet{2024Tsukui} is that we explicitly account for the inclination of each galaxy, whereas \citet{2024Tsukui} assume $i = 90^\circ$ for all systems. Projection effects can lead to significant overestimation of $z_0$ \citep{1997_degrijs}. \citet{2024Tsukui} expect a median underestimation of $\sim13\%$ in $h_r/z_0$ for single discs and up to $\sim27\%$ for thin discs due to inclination effects. Although they include these uncertainties in their reported errors of $z_0$ and $h_r/z_0$, if a significant fraction of their sample has inclinations smaller than assumed, these projection effects could still bias the median values of both quantities.
In our sample (excluding upper limits), we find a median inclination of $i = 86.1^\circ$ with a $1\sigma$ dispersion of $3.8^\circ$. A threshold of $b/a < 0.3$, corresponding to an inclination of $i > 72^\circ$ for an infinitely thin disc, is commonly adopted for the edge-on approximation \citep{Tsukui_2023}. To quantify how this assumption impacts the derived thickness, we fit the mock observations of galaxies with stellar masses of $8\times10^9~\mathrm{M_{\odot}}$ and $3\times10^{10}~\mathrm{M_{\odot}}$ (Section~\ref{sec:mockdata_gen}) generated with inclinations spanning $72^\circ - 90^\circ$, while fixing the inclination to $90^\circ$ in the fitting procedure. This leads to an overestimation of the disc thickness by up to a factor of $\approx 3$ for the more face-on cases (i.e., 72$^\circ$, see Fig.~\ref{fig:inclination_fixed}). Therefore, systematic differences between our results and those of \citet{2024Tsukui} in $z_0$ and $h_r/z_0$ may largely arise from biases associated with the assumed inclination.

\section{Implications for thin and thick disc assembly}
\label{sec:emergence}
In Section \ref{sec:results}, we show that the inferred $z_0$ and $z_0/h_r$ values are broadly consistent with those of thin discs in the Milky Way and local spiral galaxies. At the same time, their distributions exhibit extended tails toward small $z_0$ and large $h_r/z_0$ relative to local samples. This points to thin-disc assembly beginning relatively early for the bulk of the disc population. Such a trend is consistent with either (i) a thick disc that develops gradually through dynamical heating, rather than being born intrinsically thick, or (ii) a scenario in which thin and thick discs form simultaneously, but the thick-disc component is too faint in our data to be detected. 

To assess the detectability of a thick-disc component, we performed recovery tests by generating mock datasets for galaxies in our sample with measured $z_0$ (i.e., excluding the ones with upper limits on $z_0$), based on the best-fit thin disc parameters, and adding a thick disc with a range of scale lengths and scale heights, and luminosity equal to 0.1 the luminosity of the thin disc (see Appendix \ref{sec:thick} for details). We adopted this ratio of 0.1 because it matches the Milky Way and provides a conservative lower bound for local spirals, which typically have thick-disc luminosity fractions $\gtrsim 10\%$ \citep{2006_yoachim}. After PSF convolution and noise injection matched to the data, we compare the %matched-filter 
detection significance of thin-only, thick-only and thin+thick models to quantify whether such a thick disc would be detectable (see Appendix~\ref{sec:thick} for details on the significance estimation). 

These tests indicate that, even in this conservative case (a thick disc contributing only 10\% of the thin-disc luminosity), a secondary component should leave a measurable %residuals 
signature and be detectable in our data. The absence of such signals implies that any thick disc present must contribute $< 10\%$ of the thin-disc light. While this cannot be excluded observationally, there is no clear physical motivation for thick discs at high redshift to be systematically so faint: if the thick component is older, it is expected to make a relatively larger contribution at earlier times, when the thin disc is still assembling. We therefore favour scenario (i), in which the observed $z_0$ and $h_r/z_0$ values are more naturally explained by a thick disc that builds up progressively via dynamical heating.

\section{Discussion}
\label{sec:discussion}
\subsection{Is the Milky Way a special galaxy?} 

In recent years, comparisons between the properties of the Milky Way and those predicted by cosmological simulations have sparked debate over whether our Galaxy is a typical disc galaxy \citep[e.g.,][]{Evans_2020, Semenov_2024, McCluskey_2025}. Here we focus on the recent study by \citet{Xiang_2025}, who measured how the vertical-to-radial scale ratio, $z_0/h_r$, varies with stellar age in the Milky Way, extending the trend to the oldest stellar populations currently accessible. They then compared the observed relation to that of 31 Milky Way–like galaxies in the TNG50 simulations \citep{Pillepich_2019, Nelson_2019}.
The simulations indicate that stars can form in relatively thin, disc-like structures even within the first gigayear of cosmic history, but that these early discs subsequently thicken due to dynamical heating from mergers and interactions. This suggests that even the oldest, high-$\alpha$ populations in the Milky Way, characterised by $z_0/h_r \sim 1$, may have originated in an early thin disc that was later dynamically heated.
However, a quantitative offset remains: in TNG50, most stellar populations older than $\sim$10.5~Gyr have $z_0/h_r \gtrsim 0.5$, whereas in the Milky Way this is only the case for stars older than $\sim$12.5~Gyr. Only about one-sixth of the simulated analogues retain discs as thin as that of the Milky Way. \citet{Xiang_2025} suggest that this discrepancy may indicate that current simulations overestimate early disc heating, or alternatively, that the Milky Way experienced an unusually quiescent merger history. 

\begin{figure*}
    \centering
    \includegraphics[width=0.8\textwidth]{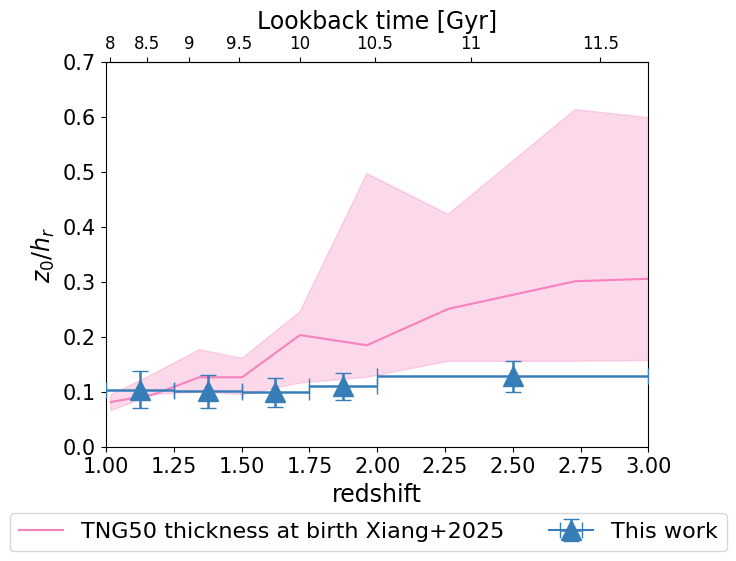}
    \caption{Ratio of scale height to scale length ($z_0/h_r$) as a function of redshift for our sample (blue triangles) and for Milky Way–like progenitors from the TNG50 simulations from \citet{Xiang_2025}. The simulation points indicate the values of $z_0/h_r$ at the time of stellar birth.}
    \label{fig:tng}
\end{figure*}

In Fig.~\ref{fig:tng}, we show $z_0/h_r$ at birth as a function of redshift as derived by \citet{Xiang_2025} from the TNG50 simulations. The blue triangles indicate the $z_0/h_r$ values from our sample (excluding upper limits). These should be regarded as upper limits to $z_0/h_r$ at birth, since these systems may have already undergone some dynamical heating (see Section~\ref{sec:caveats}). The comparison shows that simulated $z_0/h_r$ values systematically exceed those observed, suggesting that the offset between simulations and the Milky Way reflects a more general tension applicable to typical galaxies, rather than a peculiarity of the formation history of the Milky Way.

\subsection{Connecting stellar disc thickness and gas kinematics} 
As discussed in the Introduction, the formation scenarios of thick discs provide crucial constraints on the broader question of how galaxy discs form in general. One of the scenarios most commonly invoked in recent years proposes that galaxies formed their thick discs from initially thick, highly turbulent gas discs. However, support for this scenario largely comes from studies based on warm-gas tracers, even though warm gas is not the phase from which stars form.
The difficulty is that current observational constraints on disc formation and dynamics are strongly shaped by the limitations of existing facilities. Cold-gas tracers provide the cleanest measurements of disc kinematics, but observations are typically restricted to massive, highly star-forming systems at high redshift \citep[e.g.,][]{Rizzo_2021, Lelli_2021, Roman_Oliveira_2023, Tsukui_2023, 2024_rizzo}. Warm-gas tracers (e.g. H$\alpha$) are instead used to study more typical main-sequence galaxies at lower stellar masses and star-formation rates \citep[e.g.,][]{Forster_2020, Puglisi_2023, Danhaive_2025}. However, these tracers do not always reliably trace the underlying disc dynamics, as they can be affected by diffuse ionized gas as well as by inflows and outflows \citep[e.g.,][]{Kohandel_2024, Phillips_2025}.
Measurements of the stellar disc thickness provide an independent and indirect constraint on the vertical structure and kinematics of the gas from which stars formed. They therefore allow us to probe regions of the stellar-mass–SFR parameter space that remain inaccessible to cold-gas observations, and to test key assumptions commonly adopted when interpreting H$\alpha$-based kinematic measurements.

To obtain an order-of-magnitude estimate of the gas velocity dispersion from which the stars we observe, characterised by a given vertical distribution, formed, we assume that the interstellar medium (ISM) is in vertical dynamical equilibrium, such that the weight of the gas layer is balanced by the vertical momentum flux associated with turbulence. In regions where the gas self-gravity dominates over that of the other components—an assumption that is a reasonable approximation in the inner regions of $z > 1$ gas-rich, star-forming galaxies \citep{Tacconi_2020}—the gas scale height can be approximated \citep{Ostriker_2022} as
\begin{equation}
z_{0, \mathrm{gas}} \approx \frac{\sigma^2}{\pi G \Sigma_{\mathrm{gas}}},
\label{eq:hydro}
\end{equation}
where $\sigma$ is the gas velocity dispersion and $\Sigma_{\mathrm{gas}}$ is the gas surface density.
Stars formed from this gas are expected to inherit, at minimum, a comparable velocity dispersion and therefore a thickness similar to, or larger than, that of the gas layer from which they formed. To estimate the gas velocity dispersion using equation~(\ref{eq:hydro}), we assume that the gas scale height is comparable to the observed stellar scale height ($z_{0,\mathrm{gas}} \approx z_0$). We further assume that the observed stellar population formed from gas with the same surface density, accounting for stellar mass loss through the return fraction $R$ \citep{Tinsley_1980}. Under this assumption, the gas surface density can be approximated as $\Sigma_{\mathrm{gas}} = \Sigma_{\mathrm{star}}/(1-R)$, where we adopt a return fraction $R \approx 0.43$ \citep{Madau_2014}. By inverting equation~(\ref{eq:hydro}), we obtain an estimate of the gas velocity dispersion. Given the assumptions above, this estimate should be interpreted as an approximate upper limit on the velocity dispersion of the gas from which the stellar populations originated.

\begin{figure*}
    \centering
    \includegraphics[width=\linewidth]{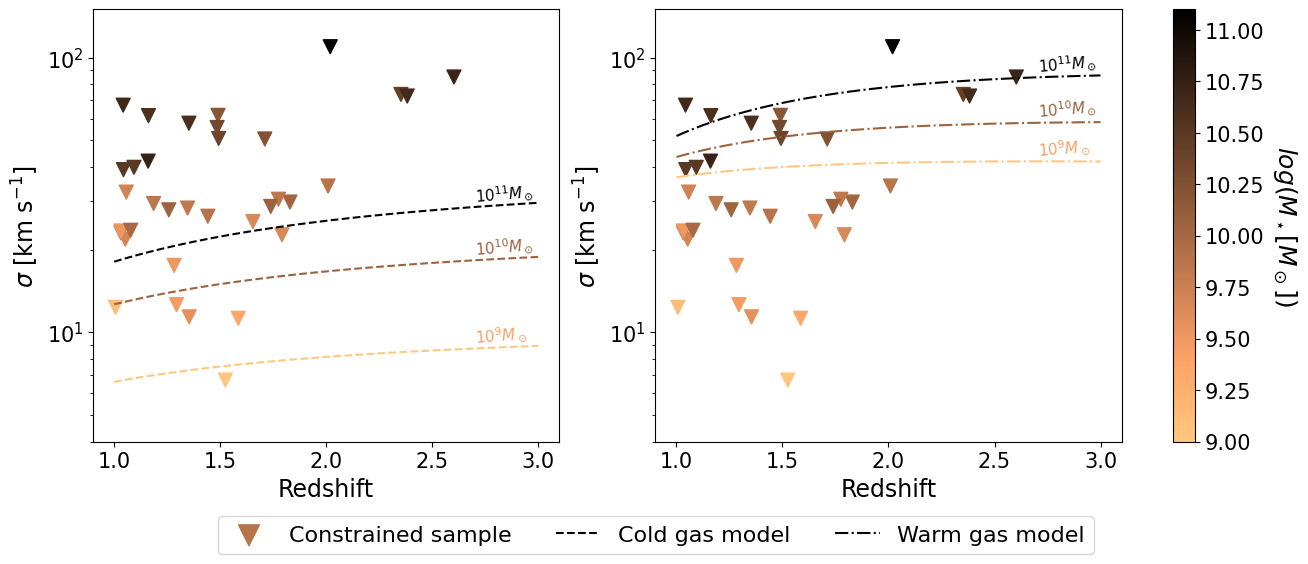}
    \caption{Upper limits on the gas velocity dispersion values, estimated from the measurements of the stellar thickness and colour-coded as a function of the stellar masses. The dashed lines show the expected evolution of the gas velocity dispersion at fixed stellar masses, as predicted by model calibrated on cold \citep[left panel,][]{2024_rizzo} and warm gas kinematics \citep[right panel,][]{Wisnioski_2025}.}
    \label{fig:sigma}
\end{figure*}

In Fig.~\ref{fig:sigma} (left panel), we present the inferred values of $\sigma$, colour-coded by stellar mass. The dashed lines show the model predictions from \citet{2024_rizzo} for the gas velocity dispersion of main-sequence galaxies as a function of redshift at fixed stellar mass. In this model, turbulence is assumed to be primarily driven by stellar feedback, and the model is calibrated on observations of highly star-forming galaxies using cold-gas tracers \citep{2024_rizzo}. Our upper limits inferred from stellar thickness are broadly consistent with these predictions, lying close to or above the model curves at fixed stellar mass. This implies that typical low-mass galaxies likely have lower velocity dispersions, on average, than the $\sim 30-40~\mathrm{km~s^{-1}}$ commonly measured in massive, highly star-forming systems via cold-gas emission lines \citep{2024_rizzo}. Although indirect, these constraints are particularly valuable for low-mass galaxies ($\lesssim 10^{10}~\mathrm{M_{\odot}}$), for which direct measurements of cold-gas kinematics will largely require next-generation facilities (e.g. ALMA2040).

In the right panel of Fig.~\ref{fig:sigma}, we compare our upper limits with the empirical model widely used to interpret H$\alpha$ kinematics \citep[e.g.,][]{2015_Wisnioski, 2018_Johnson, Danhaive_2025, Wisnioski_2025}. This model predicts systematically higher dispersions than those allowed by the constraints from stellar thickness. The resulting offset supports the interpretation that velocity dispersions measured from warm ionized gas exceed those of the cold gas phase from which stars form, consistent with a scenario in which the warm and cold ISM couple differently to stellar feedback.
In this context, H$\alpha$ velocity dispersions provide complementary information on the ionized ISM, and should be interpreted alongside other tracers when drawing conclusions about disc formation and evolution.

\subsection{Caveats} \label{sec:caveats}

In this section, we discuss the caveats and limitations of our analysis. To evaluate the vertical thickness of the galaxies in our sample, we use images at a rest-frame wavelength of $\sim1.2~\mathrm{\mu m}$. This wavelength minimizes the effects of dust attenuation and traces the bulk of the stellar population. However, this choice also means that the derived scale heights are weighted toward older stars, which have undergone dynamical heating. As a result, the inferred scale heights may be overestimated relative to those derived from filters more sensitive to younger stellar populations.

At $z=0$, studies often employ two-component disc models \citep{2006_yoachim, 2011_comeron}. For galaxies at $z>1$, however, additional tests with mock data are needed to verify whether such models can be reliably constrained given the signal-to-noise ratios and spatial resolution of current observations. A more detailed decomposition into thin and thick discs (and a bulge, where present) would provide a more complete description of the vertical structure of these systems and represents a natural avenue for future work, particularly as higher-resolution observations from next-generation telescopes (e.g., the the Extremely Large Telescope) become available. In this work, we model each galaxy with a single disc component rather than a thin+thick disc decomposition. When two discs are present, their combined light distribution can typically be approximated by a single $\mathrm{sech}^2$ profile, with the recovered $z_0$ lying between the true scale heights of the thin and thick discs, depending on their relative luminosities.
A bulge component was included for some galaxies. Although this could, in principle, obscure a compact disc, we find no systematic bias in the fitted $z_0$ or $h_r/z_0$ ratios between galaxies with and without a bulge. Our results therefore indicate the presence of a thin disc, although the derived $z_0$ values may be somewhat overestimated due to our use of a single-component model. The presence of a thick disc cannot be completely ruled out, but its surface brightness may be subdominant with respect to the thin component (see Section \ref{sec:emergence}). We note, though, that such a scenario, as well as the considerations on the adopted rest-frame wavelength discussed above do not hamper, but strengthen the results presented in this work; that is,
the evidence for the presence of thin discs as early as $z\sim2.5$.

In this work, we adopt the commonly used $\mathrm{sech}^2$ profile to describe the vertical distribution of stellar light \citep{1981_kruit}, which simplifies the integration of luminosity along the line of sight. Nonetheless, at least at $z=0$, disc flaring — the increase of the scale height with galactocentric radius, often by a factor of a few within the optical radius — is a well-established phenomenon. It has been observed in the Milky Way and in several external galaxies \citep{1982_rohlfs, degrijs1997shapegalaxydisksscale, 2002Narayan, Kasparova_2016, Mackereth_2017, Sarkar_2019, Ranaivoharimina_2024}, and is also predicted by theoretical models of disc structure \citep{Sarkar_2020, JOG_2026}. As discussed by \citet{JOG_2026}, iterative fitting of $z_0(R)$ can in principle be used to probe such radial variations. Another possible approach is to constrain the stellar scale height self-consistently from stellar kinematics under the assumption of hydrostatic equilibrium, similarly to what is done for the gas component \citep[e.g.,][]{Bacchini_2019}. However, such methodologies have seen only limited application so far, even at low redshift.
The modelling framework adopted here fits a single global value of $z_0$, and therefore cannot capture radial variations in disc thickness by construction. This represents a fundamental limitation of the present approach. Nevertheless, the methodology developed in this work already constitutes an improvement over most previous studies at high redshift, where the disc thickness has often been inferred after fixing the inclination. A more detailed characterization of disc flaring in high-redshift galaxies will likely require higher-resolution data than currently available. In particular, future observations with facilities such as MICADO on the Extremely Large Telescope will provide the spatial resolution needed to better constrain multi-parameter models of disc structure. This includes radial variations in scale height through parametric flaring profiles \citep[e.g.,][]{1982_rohlfs}, as well as constraints from stellar kinematics, both currently beyond the reach of existing facilities.

We do not account for internal dust attenuation in the disc. Because our models permit slight deviations from a perfectly edge-on orientation, one side of the disc may be slightly inclined away from the observer. The corresponding emission thus traverses a longer path length and experiences greater extinction. Although dust effects are expected to be minor at $1.2~\mathrm{\mu m}$, this asymmetry could produce a mild brightness gradient not captured by our modelling.

Finally, we note that the observed surface-brightness contours of several galaxies are less extended radially than predicted by the models. This likely reflects the presence of disc truncations, where the surface brightness declines sharply beyond a few scale lengths \citep{1981_kruit, 2011_kruit}. Such truncations are common in both thin and thick discs \citep{2012_comeron}. Incorporating a cut-off radius in future models \citep[e.g.,][]{2004_Kregel} may therefore improve the fit quality for these systems.

\section{Summary}
\label{sec:conclusion}
In this paper, we present a new methodology for measuring the thickness of edge-on disc galaxies that explicitly accounts for departures from perfectly edge-on orientations. Our approach builds a full 3D light-distribution model and uses forward modelling to incorporate PSF convolution directly in the fitting procedure. This constitutes a substantial improvement over traditional methods, which often assume an inclination of $90^\circ$, extract 1D vertical profiles, or approximate PSF effects with line-spread functions. We validate the method with mock \textit{JWST} observations and quantify the minimum disc thickness that can be robustly recovered as a function of image resolution and signal-to-noise ratio. In particular, we find that we are able to recover accurate measurements of the scale height when this is more extended than one pixel, almost independently of the signal-to-noise-ratio of the data. These tests further show that assuming galaxies to be perfectly edge-on can bias thickness measurements high by up to a factor of $\sim 2$.

We apply this model to a sample of 90 galaxies at redshifts $1 < z < 3$ with stellar masses $\gtrsim 10^9~\mathrm{M_{\odot}}$, selected from four major \textit{JWST} surveys based on data quality and low axis ratios (i.e., $\leq 0.3$). All galaxies are analysed at a rest-frame wavelength of $\sim1.2~\mathrm{\mu m}$. Of these, 46 are best fitted by a single-disc model and 44 by a disc+bulge model. For 55 galaxies, the inferred scale heights ($z_0$) are considered upper limits due to pixel-size constraints. The median scale height is $z_0 = 0.25 \pm 0.14~\mathrm{kpc}$ (excluding upper limits) and $0.26 \pm 0.09~\mathrm{kpc}$ (including them). We find no clear dependence of $z_0$ on redshift or stellar mass, though the largest $z_0$ values decrease with increasing redshift. For the ratio $h_r/z_0$, we obtain median values of $9.7 \pm 3.3$ (excluding upper limits) and $8.4 \pm 3.7$ (including them), again showing no clear trend with redshift. 

The inferred $z_0$ and $h_r/z_0$ values are broadly consistent with those of thin discs in the Milky Way and local spirals, but their distributions exhibit extended tails toward smaller $z_0$ and larger $h_r/z_0$ relative to local samples. Quantitatively, 
the median scale height of our sample is a factor of $\sim 1.6$ lower than that of local galaxies fitted with a single disc component, as adopted here.

These results can be interpreted in two ways: (i) thick discs are not born intrinsically thick, but instead build up progressively as an initially thin disc is dynamically heated over time; or (ii) thin and thick discs form simultaneously, but the thick-disc component is sufficiently faint that our single-component fits are dominated by the thin disc and any thick disc remains undetected. To investigate scenario (ii), we performed recovery tests in which we injected a thick-disc component into mock galaxies with a luminosity $L_{\rm thick}=0.1\,L_{\rm thin}$. We adopt 10\% as a conservative choice because it matches the Milky Way and lies at the low end of thick-disc luminosity fractions inferred for local spirals, which are typically $\gtrsim 10\%$. Even in this conservative case, the thick disc should produce a measurable signature and be detectable in our data, implying that any thick disc present must contribute $< 10\%$ of the thin-disc light. There is, however, no clear physical motivation for thick discs at high redshift to be systematically this faint: if anything, an older thick component would be expected to contribute relatively more at earlier times, when the thin disc is still assembling. We therefore favour scenario~(i), in which the observed $z_0$ and $h_r/z_0$ values are more naturally explained by a thick disc that develops gradually through dynamical heating. We therefore favour scenario (i), in which the observed $z_0$ and $h_r/z_0$ values are more naturally explained by a thick disc that develops gradually through dynamical heating.

Comparisons with the TNG50 simulations show that simulated galaxies typically form thicker discs than those observed in both our sample and the Milky Way. This discrepancy, previously attributed to the Milky Way’s unusually quiescent merger history, likely reflects a broader tension between simulations and observations, suggesting that real galaxies may generally form thinner discs than predicted.

Finally, measurements of stellar disc thickness provide an indirect constraint on the vertical structure and kinematics of the gas from which stars formed. Assuming vertical dynamical equilibrium, we use the observed stellar scale heights to infer upper limits on the gas velocity dispersion across stellar mass and redshift. The inferred velocity dispersions are broadly consistent with models calibrated on cold-gas observations, and indicate that normal main-sequence galaxies, currently beyond the reach of direct cold-gas kinematic measurements, likely have lower turbulence levels than the massive, highly star-forming systems for which such constraints are presently available.

\section*{Acknowledgments}
We thank the referee for a constructive report that has improved this manuscript. We thank Else Starkenburg and Cecilia Bacchini for valuable discussions and comments. FR acknowledges support from the Dutch Research Council (NWO) through the Veni grant VI.Veni.222.146. This work is based on observations made with the NASA/ESA/CSA James Webb Space Telescope. The data were obtained from the Mikulski Archive for Space Telescopes at the Space Telescope Science Institute, which is operated by the Association of Universities for Research in Astronomy, Inc., under NASA contract NAS 5-03127 for JWST. The JWST data underlying this article are publicly available and were accessed from the DAWN JWST Archive (DJA). DJA is an initiative of the Cosmic Dawn Center (DAWN), which is funded by the Danish National Research Foundation under grant DNRF140.
We acknowledge usage of the Python programming language \citep{vanrossum09}, Astropy \citep{astropycollab22}, Matplotlib \citep{hunter07}, NumPy \citep{vanderwalt11}, and Photutils \citep{photutils}.

\section*{Data Availability}

The \textit{JWST} data underlying this article are publicly available, were accessed from the DAWN \textit{JWST} Archive, and are associated with programs \#1180, \#1837.

\bibliographystyle{mnras}
\bibliography{refs}

\begin{appendix}

\section{Mock data parameters and examples}\label{appedix:mockdata}
\begin{figure*}
    \centering
    \includegraphics[width=\linewidth]{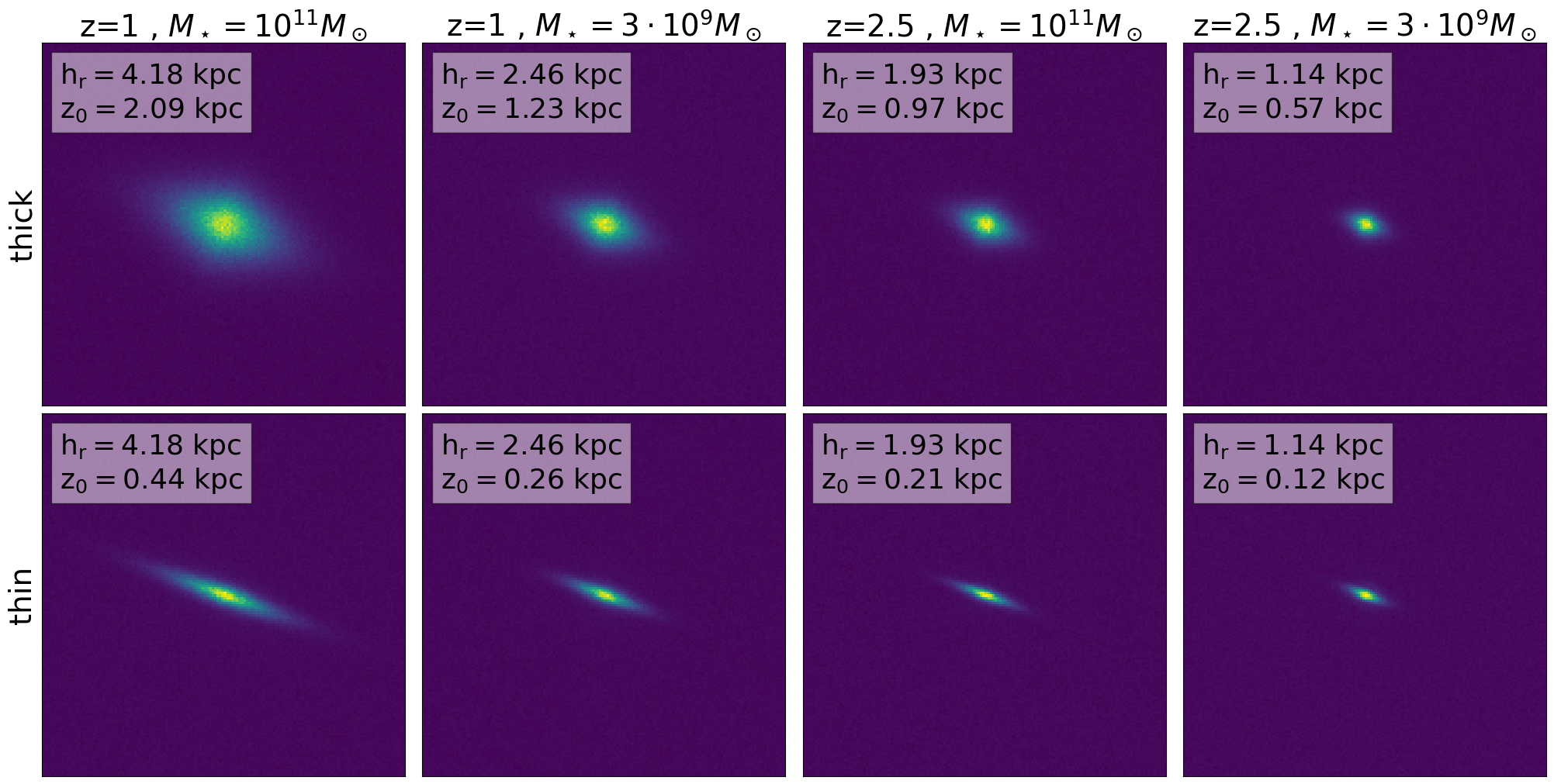}
    \caption{Representative mock-galaxy images with a thick disc (top row) and a thin disc (bottom row) at $z=1$ and $z=2.5$, shown for a low-mass case ($5 \times 10^9~\mathrm{M_{\odot}}$) and high mass case $10^{11}~\mathrm{M_{\odot}}$. All mock images shown are at high S/N.}
    \label{fig:mock_example1}
\end{figure*}

\begin{figure*}
    \centering
    \includegraphics[width=\linewidth]{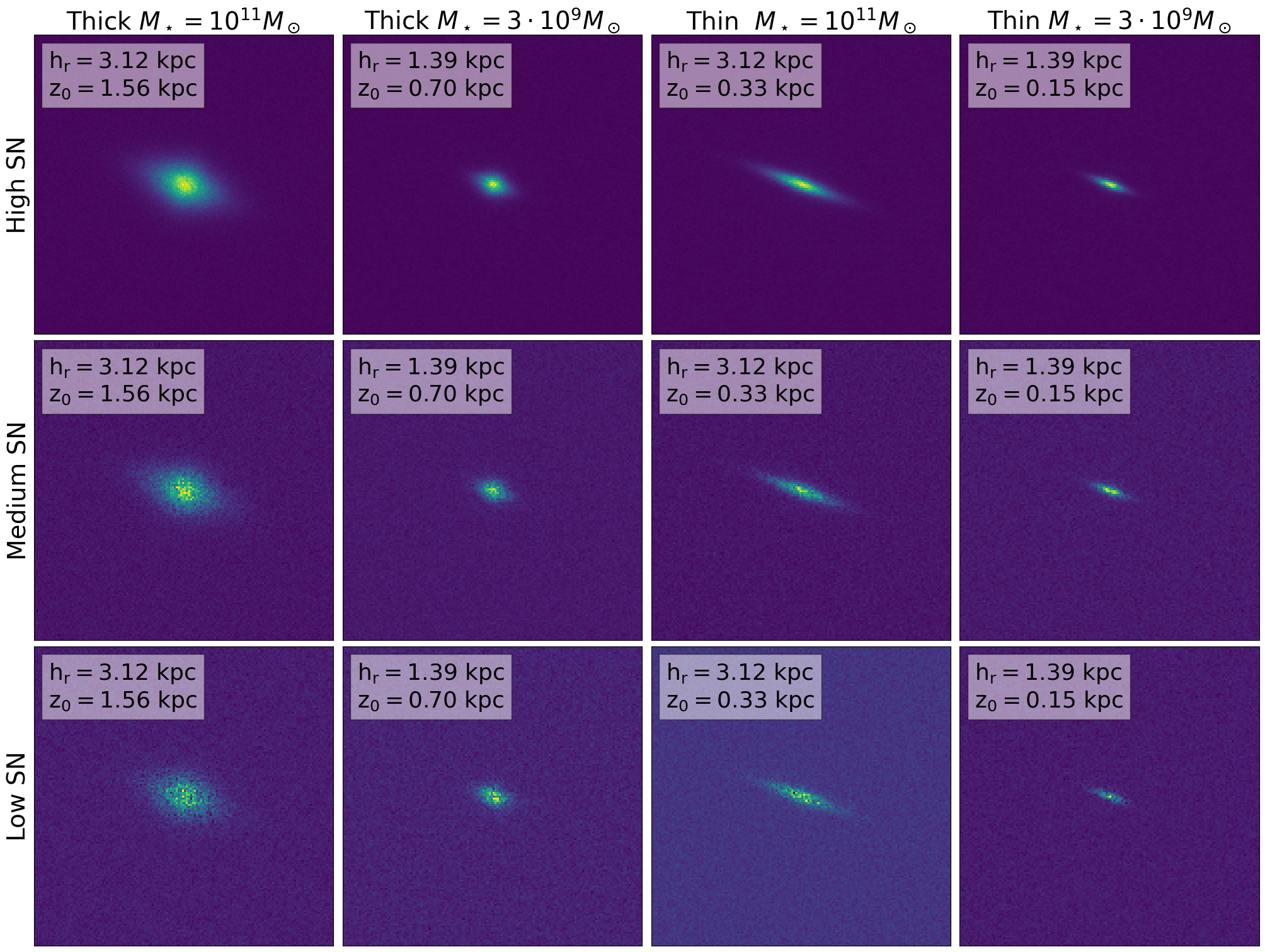}
    \caption{Representative mock-galaxy images at fixed redshift ($z=2$) shown at high, medium, and low S/N. Each column shows a galaxy generated with identical parameters (see inset), with only the S/N varied.}
    \label{fig:mock_example2}
\end{figure*}

\begin{figure*}
    \centering
    \includegraphics[width=\linewidth]{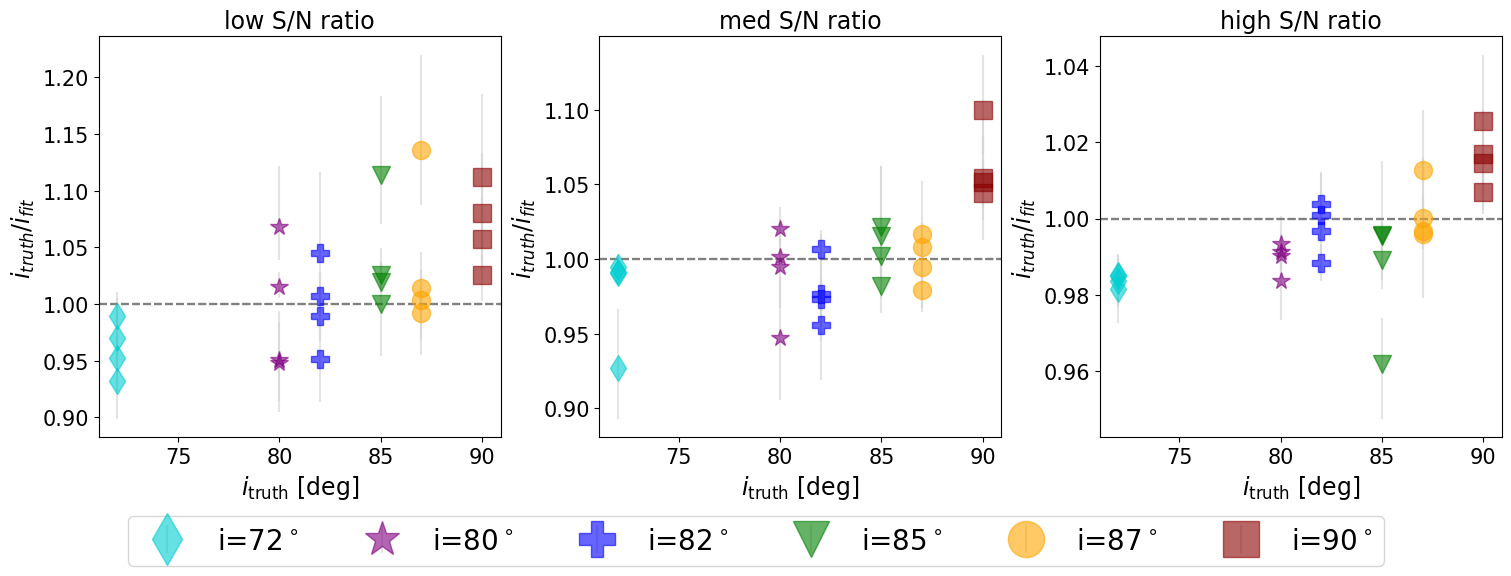}
    \caption{Recovery of the inclination $i/i_\mathrm{input}$ compared to the input. The different panels from left to right, show the recovery in a low, medium and high S/N regime. The horizontal dashed line indicates perfect accuracy.}
    \label{fig:inclination_recovery}
\end{figure*}

To validate the methodology presented in Section \ref{sec:methodology}, we generated mock data of galaxies with different sizes, scale height, geometrical parameters and S/N. In Table \ref{tab:mockdata_single}, we provide a summary of all parameters used to create the mock data and some representative examples are shown in Figs \ref{fig:mock_example1} and
 \ref{fig:mock_example2}. Table \ref{tab:error_mockdata} gives the median relative error of the recovery of the parameters $z_0$ and $h_r$. The relative error is defined as |input-fit| / input. Fig. \ref{fig:inclination_recovery} shows the recovery of the inclination $i/i_{\mathrm{input}}$ compared to the input value for a low, medium and high S/N regime. This shows that we are able to recover the inclination within a 15\% error for all S/N, and within a 5\% for high S/N data.

\begin{table*}
\begin{tabular}{llllllll}
Mass ($M_\odot$) & z   & $h_r$ (kpc) & $z_\mathrm{0,thin}$ (kpc) & $z_\mathrm{0,thick}$ (kpc) & $\theta$ ($^\circ$)                 & i ($^\circ$)                 & S/N                 \\ \hline
$3 \times 10^{9}$       & 1.0 & 1.93        & 0.21               & 0.97                & 70                             & 85                      & low, medium \& high \\
...               & 1.5 & 1.61        & 0.17               & 0.80                & ...                            & ...                     & ...                 \\
...               & 2.0 & 1.39        & 0.15               & 0.70                & ...                            & ...                     & ...                 \\
...               & 2.5 & 1.14        & 0.12               & 0.57                & ...                            & ...                     & ...                 \\ \hline
$8 \times 10^{9}$       & 1.0 & 2.4         & 0.26               & 1.2                 & ...                            & ...                     & low                 \\
...               & 1.5 & 2.00        & 0.21               & 1.00                & 70 \& {[}0,30,60,90,115,125{]}$^\mathrm{a}$ & 85 \& {[}72,80,82,87,90{]}$^\mathrm{b}$ & ...                 \\
...               & 2.0 & 1.75        & 0.19               & 0.87                & 70                             & 85                      & ...                 \\
...               & 2.5 & 1.41        & 0.15               & 0.71                & ...                            & ...                     & ...                 \\ \hline
$3 \times 10^{10}$      & 1.0 & 3.21        & 0.34               & 1.6                 & ...                            & ...                     & ...                 \\
...               & 1.5 & 2.67        & 0.28               & 1.33                & ...                            & 85 \& {[}72,80,82,87,90{]}$^\mathrm{b}$ & ...                 \\
...               & 2.0 & 2.37        & 0.25               & 1.18                & ...                            & 85                      & ...                 \\
...               & 2.5 & 1.89        & 0.20               & 0.94                & ...                            & ...                     & ...                 \\ \hline
$10^{11}$        & 1.0 & 4.18        & 0.44               & 2.09                & ...                            & ...                     & low, medium \& high \\
...               & 1.5 & 3.48        & 0.37               & 1.74                & ...                            & ...                     & ...                 \\
...               & 2.0 & 3.12        & 0.33               & 1.59                & ...                            & ...                     & ...                 \\
...               & 2.5 & 2.46        & 0.26               & 1.23                & ...                            & ...                     & ...                 \\ \hline
\multicolumn{8}{l}{\footnotesize $^\mathrm{a}$ for $i=85 ^\circ$ and low, medium \& high S/N }\\
\multicolumn{8}{l}{\footnotesize $^\mathrm{b}$ for $\theta=70 ^\circ$ and low, medium \& high S/N }
\end{tabular}
\caption{Parameters of mock data for the single disc galaxies, where the thick discs have a $h_r/z_0$ of 2.0 and thin disc have a $h_r/z_0$ of 9.4.}
\label{tab:mockdata_single}
\end{table*}

\begin{table*}
    \centering
    \begin{tabular}{ccccc}
    \hline
        S/N & $\Delta z_{0,\mathrm{thin}}/z_{0,\mathrm{thin}}$ & $\Delta z_{0,\mathrm{thick}}/z_{0,\mathrm{thick}}$& $\Delta h_{r,\mathrm{thin}}/h_{r,\mathrm{thin}}$& $\Delta h_{r,\mathrm{thick}}/h_{r,\mathrm{thick}}$ \\ \hline
        low & $0.21^{+0.27}_{-0.16}$& $0.02^{+0.04}_{-0.01}$& $0.02^{+0.02}_{-0.01}$& $0.01^{+0.02}_{-0.01}$ \\
        medium & $0.13^{+0.11}_{-0.07}$& $0.01^{+0.03}_{-0.01}$& $0.01^{+0.02}_{-0.01}$& $0.01^{+0.01}_{-0.01}$\\
        high & $0.03^{+0.04}_{-0.02}$& $0.004^{+0.004}_{-0.003}$& $0.003^{+0.003}_{-0.002}$& $0.002^{+0.003}_{-0.001}$\\ \hline
    \end{tabular}
    \caption{The mean relative error (|input-fit|/input) and and the 16\textsuperscript{th} and 84\textsuperscript{th} quantiles for $z_0$ and $h_r$ of the mock data, evaluated per S/N for thin and thick discs seperately.  }
    \label{tab:error_mockdata}
\end{table*}

\section{Detection of faint thick discs}
\label{sec:thick}
To quantify whether we would have been able to detect a thick disc with a luminosity equal to a given fraction of the thin-disc luminosity, we performed the following experiment. We restricted the analysis to galaxies with a measured value of $z_0$ (i.e. excluding upper limits). For each galaxy, we generate 10 mock datasets, each characterised by a different noise realization, using the best-fit (thin) disc parameters and then injected an additional thick-disc component. The noise is generated assuming uncorrelated gaussian distributions consistent with the reference images. 

We fixed the total (thin+thick) surface-brightness normalization at $R=0$ to the best-fit value from the real data, and assumed a thick-disc total luminosity equal to 10\% of the thin-disc luminosity. With this choice, we implicitly assume that the inner region, where the S/N is highest, is well described by the combined contribution of the thin and thick discs, and we test whether the thick-disc emission in the outskirts is sufficiently bright to be detected above the noise. A 10\% thick-to-thin luminosity fraction is consistent with the Milky Way and is a conservative assumption for local spiral galaxies, which typically show thick-disc luminosity fractions $\gtrsim 10\%$ \citep{2006_yoachim}. For the thick-disc scale length, we considered three cases: $h_{r, \rm thick} =1.6\,h_{r, \rm thin}$, consistent with the values found in local spirals \citep{2006_yoachim}, $h_{r, \rm thick} = 0.6\,h_{r, \rm thin}$, comparable with the Milky Way discs, and  $h_{r, \rm thick} = \,h_{r, \rm thin}$. For the thick-disc scale height, we explored 10 values uniformly spaced in the range $[1.6, 5.5]\,z_0$ of the thin disc, consistent with the typical values found in local spirals \citep{2006_yoachim}. Each mock image was convolved with the \textit{JWST} PSF and Gaussian noise was added to match the rms of the real data.

To quantify the detectability of the thick-disc component, we evaluate the maximum-likelihood detection significance, equivalent to the case of a matched filter experiment with matching kernel equal to the reference model itself. In such a case, the detection significance $\mathrm{DS}$ is defined as (see, e.g., \citealt{Zulbeldia_2019} for a review),

\begin{equation}
\mathrm{DS} = 
\frac{\mathbf{m}^{T}\mathbf{C}^{-1}\mathbf{d}}
{\left[\mathbf{m}^{T}\mathbf{C}^{-1}\mathbf{m}\right]^{1/2}}
\end{equation}
where $\mathbf{m}$ is the PSF-convolved noiseless model and $\mathbf{d}$ is the corresponding noisy mock dataset, $\mathbf{C}$ is the covariance matrix, assumed to be diagonal, with variances set by the rms noise of the observations. The significance reported throughout this work is computed as the average of the $\mathrm{DS}$ estimates from the different noise realizations.

\begin{figure*}
    \centering
    \includegraphics[width=\linewidth]{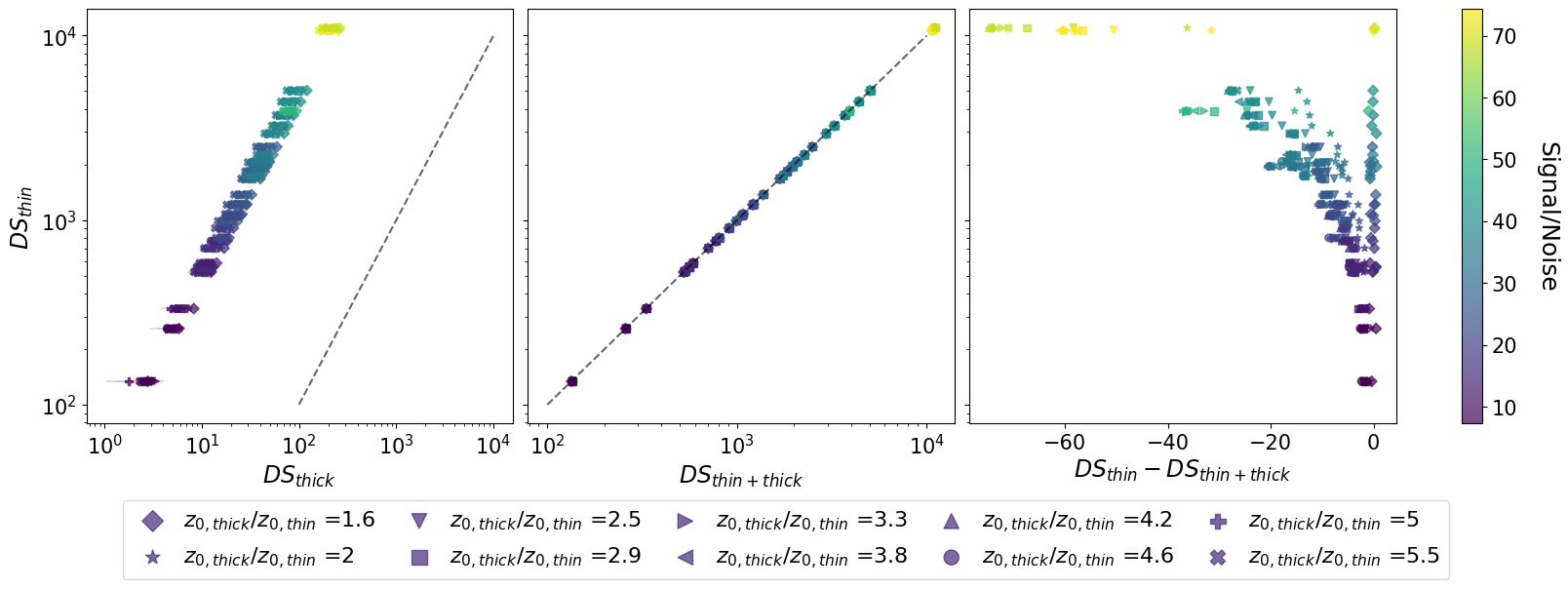}
    \caption{Detection significance of the single thin disc compared to a double disc model, with $h_{r,\mathrm{thick}}=1.6 \, h_{r,\mathrm{thin}}$. The left panel shows $DS_\mathrm{thick}$ vs  $DS_\mathrm{thin}$, the middle panel show the $DS_\mathrm{thin+thick}$ compared to $DS_\mathrm{thin}$, and the right panel shows the difference $DS_\mathrm{thin}$-$DS_\mathrm{thin+thick}$ vs $DS_\mathrm{thin}$. The different marker indicate the ratio $z_\mathrm{0,thick}/z_\mathrm{0,thin}$ and are coloured according to the S/N of the data. The dashed line shows the 1:1 ratio.}
    \label{fig:DS_discs}
\end{figure*}

\begin{figure*}
    \centering
    \includegraphics[width=\linewidth]{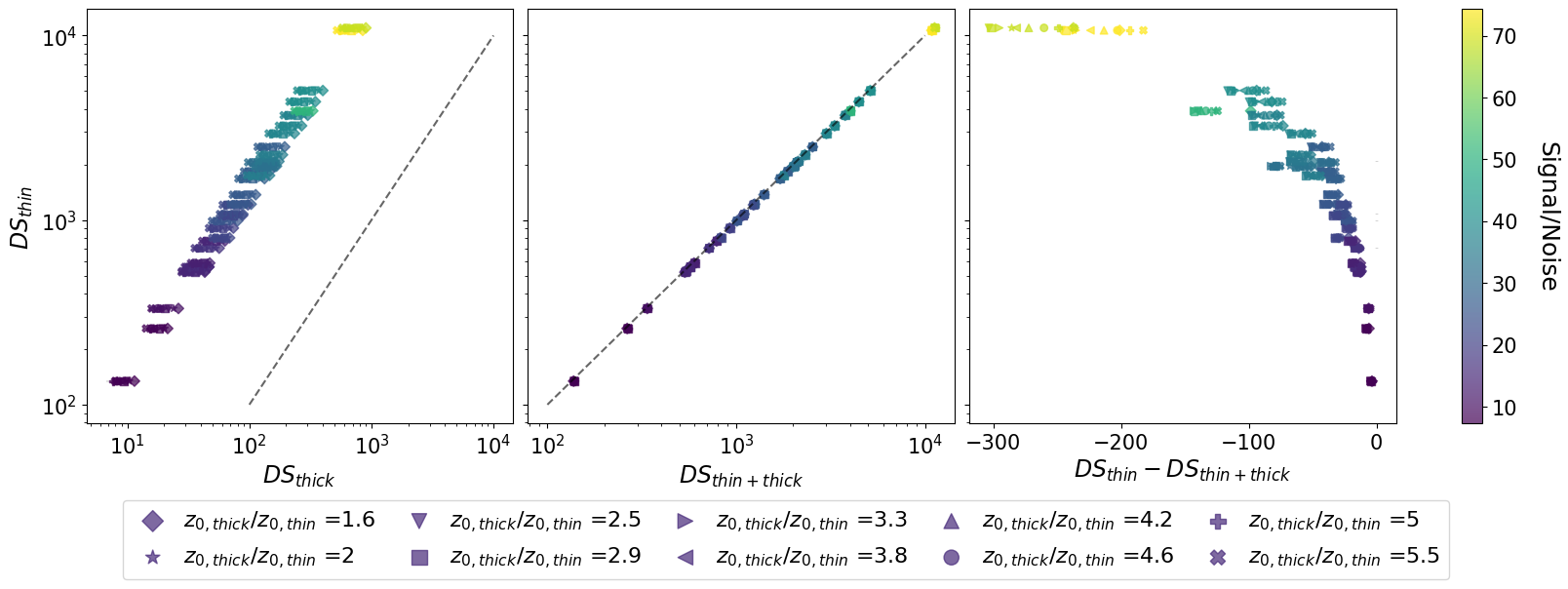}
    \caption{Same as Fig. \ref{fig:DS_discs} but with $h_{r,\mathrm{thick}}=h_{r,\mathrm{thin}}$ }
    \label{fig:DS_discs_same_extend}
\end{figure*}

\begin{figure*}
    \centering
    \includegraphics[width=\linewidth]{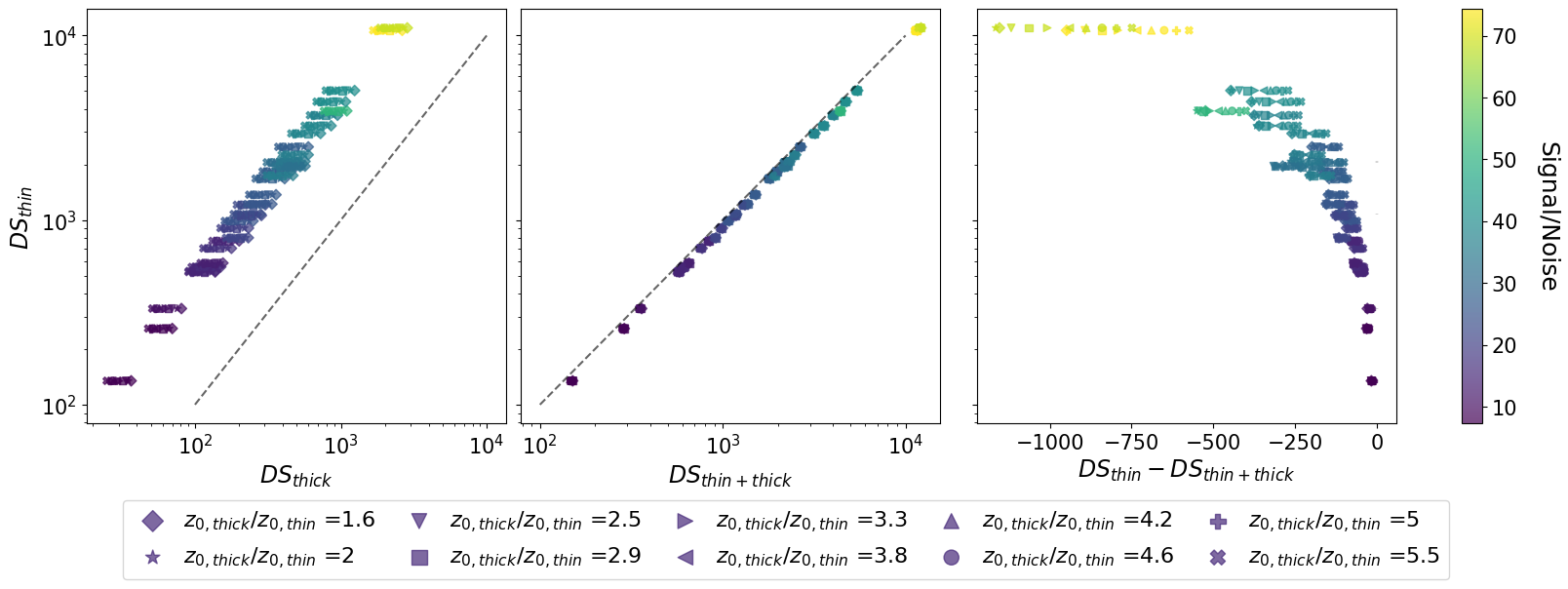}
    \caption{Same as Fig. \ref{fig:DS_discs} but with $h_{r,\mathrm{thick}}= 0.6\, h_{r,\mathrm{thin}}$ }
    \label{fig:DS_discs_extended_thin}
\end{figure*}

For each galaxy and each set of mock realizations, we compute $\mathrm{DS}$ for thin-disc-only, thick-disc-only model and a thin+thick disc model. In Fig. \ref{fig:DS_discs}, we show the values of $\mathrm{DS}$ for the sets of 10 mock data for each galaxy of the sample for galaxies with $h_{r, \rm thick} =1.6\,h_{r, \rm thin}$. The corresponding figures for the cases $h_{r, \rm thick} =1.6\,h_{r, \rm thin}$ and $h_{r, \rm thick} = h_{r, \rm thin}$ are shown in Figs \ref{fig:DS_discs_same_extend} and \ref{fig:DS_discs_extended_thin}. The markers are colour-coded according to the S/N of the reference observation (computed as described in Section \ref{sec:data}). The comparison between the thin-disc-only and thick-disc-only models (left panel of Figs~\ref{fig:DS_discs}--\ref{fig:DS_discs_extended_thin}) shows that, although faint, a thick-disc component remains distinguishable from the background noise ($\mathrm{DS}=1$). However, for the two lowest--S/N galaxies (S/N $< 10$) the expected thick-disc signal is likely too weak to reach a significance level that would allow a confident detection.

The comparison between the $\mathrm{DS}$ for the thin-disc-only model and the thin+thick disc model (middle panel in Figs \ref{fig:DS_discs}--\ref{fig:DS_discs_extended_thin}) suggests that the two model have a similar detection significance. The right panel in Figs \ref{fig:DS_discs}--\ref{fig:DS_discs_extended_thin} gives the residuals $\mathrm{DS}_\mathrm{thin}- \mathrm{DS}_\mathrm{thin+thick}$. For most mock data the residuals left by the thick disc component have sufficient DS to be considered a detection ($\mathrm{DS}\gtrsim5$). For galaxies with $h_{r, \rm thick} =1.6\,h_{r, \rm thin}$, observations with low S/N data or a ratio $z_{0,\mathrm{thick}}/z_{0,\mathrm{thin}} < 2$ the thick disc is not considered detectable. We note that, in the latter regime, the secondary component would be more appropriately interpreted as an additional thin-disc component rather than a classical thick disc. For galaxies with $h_{r, \rm thick} \leq\,h_{r, \rm thin}$, the thick disc is not detectable in observations observations with low S/N data (S/N $< 10$) or S/N $\approx 10$ when $z_{0,\mathrm{thick}}/z_{0,\mathrm{thin}} < 2$.

\end{appendix}

\bsp
\label{lastpage}
\end{document}